\documentclass[letterpaper,11pt]{article}

\usepackage{jheppub} 

\usepackage{dsfont}

\usepackage{subfig}
\usepackage{float}

\def\ben{\begin{equation}}
\def\een{\end{equation}}
\def\bea{\begin{eqnarray}}
\def\eea{\end{eqnarray}}
\def\vx{{\vec{x}}}
\def\cO{{\cal{O}}}
\def\dt{\delta t}

\def\nn{\nonumber}

\title{Quantum Quench in Non-relativistic Fermionic Field Theory: Harmonic traps and 2d String Theory}

\author{Sumit R. Das,}
\author{Shaun Hampton,}
\author{Sinong Liu.}

\affiliation{Department of Physics and Astronomy, University of Kentucky, Lexington, KY 40506, U.S.A.}

\emailAdd{das@pa.uky.edu}
\emailAdd{sha444@uky.edu}
\emailAdd{sinong.liu@uky.edu}

\abstract{We investigate a class of exactly solvable quantum quench
  protocols with a finite quench rate in systems of one dimensional non-relativistic fermions in external harmonic
  oscillator or inverted harmonic oscillator potentials, with time
  dependent masses and frequencies. These hamiltonians arise, respectively,
 in harmonic traps, and the $c=1$ Matrix
  Model description of two dimensional string theory with time
  dependent string coupling. We show how the dynamics is determined by a single function of time which satisfies a generalized Ermakov-Pinney equation. The quench protocols we consider asymptote to constant masses and frequencies at early times, and cross or approach a gapless potential. In a right side up harmonic oscillator potential we determine the scaling behavior of the one point function and the entanglement entropy of a subregion by obtaining analytic approximations to the exact answers. The results are consistent with Kibble-Zurek scaling for slow quenches and with perturbation calculations for fast quenches. For cis-critical quench protocols the entanglement entropy oscillates at late times around its initial value. For end-critical protocols the entanglement entropy monotonically goes to zero inversely with time, reflecting the spread of fermions over the entire line. For the inverted harmonic oscillator potential, the dual collective field description is a scalar field in a time dependent metric and dilaton background.}

\begin{document}

\begin{flushright}
{UK/19-02}
\end{flushright}

\maketitle
\flushbottom

\section{Introduction}
\label{sec:intro}
A common way to study non-equilibrium properties of quantum
field theories is to subject them to a quantum quench,
i.e. introduce an explicit time dependence to parameters which appear
in the lagrangian. Among other things, this is interesting for several
reasons. One motivation is to study equilibration and possible
thermalization of these systems. Suppose the time dependent parameters approach constant values in the far
past and future, and the system is initially in the ground state. The
quench then excites the system. At late times,
when the parameter again becomes a constant (which is generally
different from the initial value), one would like to know the nature
of the excited state, and if it is approximately described by a
thermal state in some appropriate sense. 

A second motivation - which is one of our main interests - is to
study dynamics in critical phase transitions \cite{senguptarmp}. Suppose the initial
hamiltonian is gapped, while the quench protocol crosses or approaches
a critical point where the gap vanishes. On general grounds one
expects that various observables would obey universal behavior. 

An early example of such a universal behavior is Kibble Zurek scaling for global quenches \cite{kibblezurek, qcritkz} (where the parameters depend only on time). This holds in many systems when the time scale of the quench $\dt$ is large compared to the inverse of the initial energy gap $E_g$. In this case, the initial time evolution is adiabatic. However since the instantaneous gap is descreasing with time, adiabaticity breaks down at some time called the Kibble Zurek time $t_{KZ}$.  This is typically determined by the Landau criterion,
\ben
\frac{1}{E_{gap}(t)^2}\frac{dE_{gap}(t)}{dt} \vert_{t= t_{KZ}} \sim 1
\label{0-1}
\een
where $E_{gap}(t)$ is the instantaneous energy gap. This equation then determines $t_{KZ}$ in terms of $\dt$.
According to the assumptions of Kibble and Zurek, the system soon enters a {\em diabatic} regime, and the instantaneous correlation length at the Kibble Zurek time is the only length scale in the problem. 

In the following we will follow standard nomenclature to distinguish several classes of quench protocols. The first protocol is called a trans-critical protocol (TCP). Here the system begins in a gapped phase and the coupling varies monotonically across a critical value, and approaches a final value which also corresponds to a gapped phase.  The second is called a cis-critical protocol (CCP) where the time dependence is not monotonic. Here the system starts from a gapped phase, approaches a critical point and reverts back to a constant value which also corresponds to a gapped phase. The third protocol is called a end-critical protocol (ECP). Here the system begins in a gapped phase and monotonically approaches a critical point at infinitely late time. In TCP or CCP, the response at early times then scales as appropriate power of the correlation length, leading to a scaling as some universal power of the quench time scale. For example the one point function of an operator $\cO$ will scale as
\ben
\langle\cO\rangle \sim \xi_{KZ}^{-\Delta}
\een
where $\Delta$ is the dimension of $\cO$. 
For ECP, the appropriate scaling variable is the energy scale at the Kibble Zurek time. At any given time the response will be adiabatic for sufficiently large $\dt$, while for a small enough $\dt$ there will be a Kibble Zurek regime \cite{entkz}.

Even though these assumptions appear to be drastic, such a scaling - together with an accompanying mechanism for defect formation in symmetry breaking transitions - appears to hold for many systems. 
Kibble Zurek scaling has been studied in a variety of solvable models and in holographic setups \cite{kzholo}. The latter provide some insight into the origins of universality. The best known results involve one point functions (e.g. defect density) and correlation functions. However similar scaling holds for the entanglement entropy of a subregion in some model $1+1$ dimensional systems. 

At the other extreme is {\em instantaneous} quench where a sudden change of a parameter causes the system to go from a gapped phase to a critical point abruptly. In this case, universal results are known for correlators and entanglement entropies of $1+1$ dimensional systems \cite{cc2,cc3}.  Of particular interest is the spread of entanglement with time \cite{cc2} - this kind of spread has been conjectured to hold for higher dimensional systems \cite{tsunami} and there has been evidence for this in holographic calculations \cite{entholo} as well as in free field theories. 

More recently it has been found that in a relativistic theory there is an intermediate regime where a different universal scaling holds \cite{dgm1}- a result which was first found in holographic calculations \cite{numer,fastQ} and later found to hold quite generally. Consider a relativistic quantum field theory in $d$ dimensional space-time which is obtained by the RG flow from a UV fixed point. The action can be then written as
\ben
S = S_{CFT} - \int dt \int d^{d-1}x ~\lambda (t) \cO_\Delta (\vx,t)
\label{0-2}
\een
Here $S_{CFT}$ stands for the conformal field theory action at  the UV fixed point and $\Delta$ denotes the conformal dimension of the operator $\cO_\Delta (\vx,t)$ in this CFT. The time dependent coupling $\lambda(t)$ goes from a constant value $\lambda_0$ in the infinite past some other value $\lambda_1$ in the distant future, and the time dependence is in some time interval of size $\dt$. Then this regime is defined by
\ben
\Lambda_{UV}^{-1} \ll \dt \ll (\delta \lambda)^{-\frac{1}{d-\Delta}}, 
(\lambda_{0,\pm})^{-\frac{1}{d-\Delta}}
\label{0-3}
\een
where $\lambda_\pm$ denote the largest and smallest value of the coupling and $\delta \lambda$ is the excursion of the coupling during the quench process. In this regime the one point function soon after the quench is over scales as
\ben
\langle\cO_\Delta\rangle \sim (\dt)^{d-2\Delta}
\label{0-4}
\een
This is a result in {\em any} relativistic field theory, and follows from two basic properties \cite{dgm1}. The first is causality. The second property is that the causal Green's functions of a massive theory become those of the UV conformal theory for space-time separations which are small compared to the inverse mass gap. Once these properties hold, it turns out that the dimensionless parameter which controls time dependent perturbation theory is the combination of the coupling with an appropriate power of $\dt$, and all other scales go away. This combination is small in the fast quench limit and the result (\ref{0-4}) follows from the lowest order perturbation theory. This regime of scaling has been investigated explicitly in free field theories with time dependent masses and in conformal field theories with relevant and marginal deformations \cite{smolkin}. In continuum free theories there appears to be a smooth transition between Kibble-Zurek and Fast scaling regimes \cite{dgm3}, while in lattice theories this connects to the abrupt quench regime at quench rates at the scale of the lattice spacing \cite{ddgms}. Apart from one point functions, the whole range of scaling behavior is visible in quantities like the entanglement entropy \cite{dascaputa} as well as circuit complexity \cite{complexscaling} \footnote{Other aspects of time dependence of complexity following a quench have been studied earlier in \cite{othercomplex}.}.

In many situations, particularly in experimental setups, one is interested in non-relativistic systems. Our ultimate goal is to investigate whether there are universal scaling laws which hold in non-relativistic systems. While Kibble Zurek scaling is expected to hold, the status of fast quench scaling is unclear. In specific models where non-relativistic Lifshitz type dispersion relations appear, e.g. the anisotropic critical points of the Kitaev model one indeed finds fast quench scaling with appropriate scaling dimensions \cite{ddgms}. More generally, Lieb Robinson bounds \cite{liebrobinson} for lattice non-relativistic systems may provide the necessary ingredient. Indeed in recent work in lattice models with dynamical exponent $z \neq 1$ it has been found that the spread of entanglement following a sudden quench indeed has an effective finite velocity \cite{ali}. However, such a finite speed has been also observed in non-relativistic systems which do not obey Lieb-Robinson bound \cite{new}.

In this work we study the issue of scaling in a specific solvable system : a system of $N$ mutually non-interacting non-relativistic fermions in a harmonic or inverted harmonic potential with a time dependent frequency and a time dependent mass. Using the results of \cite{pinney} will show how the problem of quantum quench with some {\em smooth quench profile} in such systems can be solved analytically once one can solve a nonlinear equation (Ermakov-Pinney (EP) equation). The solutions of this equation can be in turn determined in terms of the solutions of the {\em classical} equation of motion of a single particle in the same harmonic potential.

Indeed harmonic traps are of considerable interest in experimental cold atom physics : quantum quench experiments often involve release of particles from harmonic traps. 

Our interest in the inverted harmonic oscillator potential on the other hand stems from its connection to two dimensional string theory \cite{cone}. As is well known, the double scaled limit of the singlet sector of the quantum mechanics of a single hermitian matrix reduces to a set of fermions in an inverted harmonic oscillator potential. The string coupling appears as the mass of the fermion. Thus two dimensional string theory with a time dependent coupling reduces to the problem of fermions with time dependent mass in such a potential \footnote{Fermions in harmonic oscillator potentials also appear in the description of special states in the AdS/CFT correspondence \cite{llm,mandalhalf}. Introducing a time dependent mass for such fermions naively corresponds to a time dependent coupling of the Yang-Mills theory. However this breaks supersymmetries : the truncation of matrix models and therefore fermions do not hold any more.}. String theory with time dependent string couplings have been studied extensively in the context of AdS/CFT to investigate thermalization via black hole formation \cite{chesler}. In a different context these have been used as models of AdS cosmology \cite{dascosmo,chuho,turok,horocosmo,dasbrand}, but the outcome has been rather inconclusive.  Here we hope to obtain exact results in a simplified situation. 

In this paper we will set up the formalism necessary to solve both the harmonic and inverted harmonic potential problems. We present detailed results for the problem in harmonic trap : the problem of two dimensional string theory will appear in a future publication \cite{future}.

We will solve the quantum mechanical time evolution of such a system for interesting time dependent frequencies of the CCP and ECP type and calculate the early time response of one point functions as well as entanglement entropies for a sub-region for arbitary quench rates to find the scaling behavior in various regimes. We will also explore the late time behavior of the entanglement entropy. We find Kibble-Zurek scaling for slow quenches, while for fast quenches we show that the result scales in a way which is consistent with time dependent perturbation theory. At late times the entanglement entropy in a CCP oscillates with an amplitude which appears to remain constant in time. This reflects the lack of thermalization of the system. For the ECP the entanglement entropy monotonically goes to zero as a power law in time, reflecting the fact that the particles can now spread all over space.

Such solvable systems have played a major role in providing insight into scaling properties of quantum quench in continuum relativistic theories and in spin systems which can be reduced to lattice versions of relativistic fermions \cite{dgm1,dgm3,ddgms}. As we will see, our example may not be the appropriate setup to explore a possible universal scaling at fast rates. Neverthless, we hope that these exact solutions will provide some insight into the general problem.

Abrupt quantum quench in a system of free non-relativistic fermions which arise from Matrix Quantum Mechanics with various potentials has been investigated in several papers \cite{mandal,calabrese,ruggiero,minguzzi,scopa2}. In particular \cite{mandal} has extensively studied the problem in terms of the dynamics of the Wigner phase space density, investigated approach to a generalized Gibbs ensemble and discovered interesting dynamical phase transitions. The papers \cite{calabrese,minguzzi,scopa2} deal with the fermion problem directly in the presence of various kinds of abrupt quenches. Other aspects of the dynamics in such fermion systems (e.g. shock wave formation) have been studied in \cite{kulkarni}.

The paper \cite{ruggiero} considers the dynamics of the Wigner phase space density as well as a system of bosons and fermions using methods similar to us, in particular the EP equation. The paper \cite{scopa} considers slow smooth quenches for bosons also using the EP equation. The EP equation has also been used to study entanglement dynamics following an abrupt quench in a harmonic chain in \cite{ghosh}.

Our work is complementary to these papers. We are interested in studying scaling of various quantities as functions of the quench rate. We have been able to find exact analytic solutions to several smooth quench protocols which we use for this purpose.

In section \ref{sec-two} we set up the second quantized fermion field theory and show how this can be solved exactly for $\pm x^2$ potentials in terms of a function $\rho(t)$ which satisfies generalized Ermakov-Pinney equation and show how to obtain its solutions. In section \ref{sec-three} we quantize these theories in the Heinsenberg picture "in" state and show how observables can be expressed entirely in terms of $\rho(t)$. In section \ref{sec-five} we provide exact solutions for some CCP and ECP quench protocols for the harmonic problem. Sections \ref{sec-six} - \ref{sec-eight} contain our results for the one point function of the quenched operator and the entanglement entropy for these protocols and their scaling as functions of the quench rate. Section \ref{sec-nine} deals with comments about the behavior of the phase space density.

\section{Fermion field theory}
\label{sec-two}

Consider a system of $N$ non-relativistic fermions in $1+1$ dimensions with a hamiltonian given by
\ben
H = \int dx~\psi^\dagger(x)\left[-\frac{\hbar}{2 m(t)} \frac{\partial^2}{\partial x^2} \pm \frac{1}{2 \hbar} m(t) \nu^2(t) x^2 \right] \psi (x)
\label{1-1}
\een
where $m(t), \nu (t)$ are real smooth functions. The Schrodinger picture fermion field operators above satisfy the usual anti-commutation relations
\bea
\{ \psi (x) , \psi^\dagger (x') \} & = &  \delta (x-x') \nn \\
\{ \psi (x) , \psi (x') \} & = & \{ \psi^\dagger (x) , \psi^\dagger (x') \} = 0
\label{1-1dd}
\eea
The condition that the total number of fermions is $N$ then leads to the constraint
\ben
\int_{-\infty}^\infty dx~ \psi^\dagger (x,t) \psi(x,t) = N
\label{1-1c}
\een
The plus sign in (\ref{1-1}) is the hamiltonian of particles with a time dependent mass in a harmonic trap with a time dependent frequency. The minus sign with $\nu = 1$ is the hamitonian of the singlet sector of the double scaled  single hermitian matrix quantum mechanics which is dual to two dimensional string theory with a time dependent string coupling $g_s(t) = m(t)$. The Heisenberg picture equation of motion is the Schrodinger equation
\ben
i \frac{\partial \psi (x,t)}{\partial t} = \left[-\frac{\hbar}{2 m(t)} \frac{\partial^2}{\partial x^2} \pm \frac{1}{2\hbar} m(t) \nu^2(t) x^2 \right] \psi(x,t)
\label{1-3}
\een
In the following we will set $\hbar = 1$.

\subsection{The general solution}

In terms of a new time variable $\tau$
\ben
d\tau = \frac{dt}{m(t)}
\label{1-4}
\een
we can transfer the time dependence of the mass to the frequency term and (\ref{1-3}) becomes
\ben
i \frac{\partial \psi (x,\tau)}{\partial \tau} =
\left[-\frac{1}{2}\frac{\partial^2}{\partial x^2} \pm \frac{1}{2} \omega^2(\tau) x^2 \right] \psi (x,\tau)
\label{1-5}
\een
where 
\ben
\omega (\tau) = m(t) \nu (t)
\label{1-6}
\een
A solution to the equation (\ref{1-5}) can be obtained in terms of the solution of the Schrodinger equation with a constant mass and a constant frequency as follows \cite{pinney}. First define a new field $\Phi (x,\tau)$ by
\ben
\psi (x,\tau) = {\rm exp} [ - \alpha (\tau) x^2 - \beta (\tau) ]~\Phi (x,\tau) 
\label{1-7}
\een
Secondly, make a change of variables 
\bea
\tau \rightarrow T & = & \int^\tau \frac{d\tau'}{\rho (\tau')^2} \nn \\
x \rightarrow y & = & \frac{x}{\rho(\tau)}
\label{1-8}
\eea
Then $\Phi(y,T)$ satisfies
\ben
i \frac{\partial \Phi (y,T)}{\partial T} =
\left[-\frac{1}{2}\frac{\partial^2}{\partial y^2} \pm \frac{1}{2} y^2 \right] \Phi (y,T)
\label{1-9}
\een
provided 
\ben
\beta (\tau) = \log [\rho (\tau)]^{1/2}~~~~~~\alpha (\tau) = - i \partial_\tau \beta (\tau)
\label{1-10}
\een
and the function $\rho (\tau)$ satisfies a generalization of the Ermakov-Pinney equation \cite{pinneyeqn}
\ben
\partial_\tau^2 \rho (\tau) \pm \omega(\tau)^2 \rho (\tau) = \pm \frac{1}{\rho(\tau)^3}
\label{1-11}
\een
Here the positive sign refers to the right-side up harmonic oscillator while the negative sign refers to the inverted harmonic oscillator of relevance to the hermitian matrix model. The latter case will be discussed in detail in \cite{future}.

In the adiabatic approximation the function $\rho (\tau)$ is simply $\frac{1}{\sqrt{\omega(\tau)}}$. A departure from this value denotes a departure from adiabaticity and describes the exact response.

Furthermore the most general solution of  (\ref{1-11}) is given by
\ben
\rho (\tau)^2 = A f(\tau)^2 + 2B f(\tau)g(\tau) + C g(\tau)^2
\label{1-12}
\een
where $A,B,C$ are constants and $f(\tau), g(\tau)$ are two linearly independent solutions of the {\em classical} equation of motion of a single particle moving in a harmonic (inverted harmonic) potential with the same time dependent frequency $\omega (\tau)$
\ben
\partial_\tau^2 X \pm \omega(\tau)^2 X = 0
\label{1-13}
\een
Furthermore $A,B,C$ must satisfy
\ben
AC - B^2 = \pm \frac{1}{Wr(f,g)^2}
\label{1-14}
\een
where $Wr(f,g) = f\partial_\tau g - g \partial_\tau f$ is the wronskian of the two solutions. By the equations of motion this is a constant in time and can be therefore evaluated at any time.

The problem of fermions with a time dependent mass in a harmonic (or inverted harmonic) potential with a time dependent frequency can be therefore reduced to a problem with a constant mass and a constant frequency. The only equation one needs to solve is the classical equation (\ref{1-13}). As we will see below, many quantities of physical interest can be expressed entirely in terms of the function $\rho (t)=\rho(\tau)$.

A general solution of the equation (\ref{1-9}) has the form
\ben
\Phi_n(y,T) = N_n~e^{-i\lambda(n)T} \phi_n(y)
\label{1-14-1}
\een 
where $\phi_n$ denote a complete orthonormal set of eigenfunctions of the hamiltonian given by the right hand side of (\ref{1-9}) with eigenvalue $f(n)$. Then the above discussion implies that a general solution of the equation (\ref{1-5}) may be written as
\ben
\psi_n(x,\tau) =\frac{1}{\sqrt{\rho(\tau)}}~{\rm exp}\left[\frac{i}{2}\frac{\partial_\tau \rho(\tau)}{\rho(\tau)} x^2\right]~\Phi_n \left(\frac{x}{\rho(\tau)},T\right)
\label{1-14-2}
\een
The orthonormality conditions for the eigenfunctions $\phi_n(y)$ then imply the orthonormality conditions for the solution (\ref{1-14-2}).

The form of the solution (\ref{1-14-2}) reveals another physical meaning for the function $\rho(\tau)$. In the wavefunctions of a harmonic oscillator at fixed frequency $\omega$, one can rescale out the frequency by $x \rightarrow \sqrt{\omega} x$ and $\tau \rightarrow \omega \tau$. The normalization of the wavefunction also involves $\omega^{1/4}$, as would be required by the rescaling of $x$. In our problem $\rho (\tau)$ {\em almost} plays the role of such a time dependent rescaling. The term which spoils this is the phase factor which involves $\partial_\tau \rho(\tau)$. This is of course consistent with the fact that in lowest order of adiabatic approximation $\rho (\tau) = \frac{1}{\sqrt{\omega(\tau)}}$.

The function $\rho(\tau)$ is given by (\ref{1-12}). The independent solutions $f(\tau),g(\tau)$ and the constants $A,B,C$ have to be chosen so that the solutions (\ref{1-14-2}) satisfy the correct initial condition.

\subsection{Solution in terms of Phase Space Density}
\label{sec-phasespace}

It will be useful to think in terms of the Wigner phase space density operator
\ben
u(q,p,t) = \int dx e^{ipx/\hbar} \psi^\dagger (q-x/2) \psi(q+x/2)
\label{1-15}
\een
The condition (\ref{1-1dd}) then becomes
\ben
u(q,p,t)\star u(q,p,t) = u(q,p,t)
\label{1-15b}
\een
while (\ref{1-1c}) becomes
\ben
\int \frac{dqdp}{2\pi \hbar}u(q,p,t) = N
\label{1-15a}
\een
where $\star$ denotes the Moyal star product. As shown in \cite{dhar-mandal-wadia} the fermion field theory can be expressed as a path integral in terms of these variables with a co-adjoint orbit action. Formulating the theory in terms of $u(p,q,t)$ is particularly useful in the classical limit $\hbar \rightarrow 0, N \rightarrow \infty$ with $N\hbar$ held fixed. In this limit the Moyal product reduces to an ordinary product.  

In this limit the operator $u(p,q,t)$ satisfies the equation
\ben
[\partial_\tau + p\partial_q \mp \omega^2(\tau) q\partial_p]u(p,q,\tau) = 0
\label{1-16}
\een
If we make the change of variables 
\bea
\tau \rightarrow T & = & \int^\tau d \tau' \frac{1}{\rho (\tau')^2} \nn \\
q \rightarrow Q & = & \frac{q}{\rho(\tau)} \nn \\
p \rightarrow P & = & p \rho(\tau) - q \partial_\tau \rho (\tau) 
\label{1-17}
\eea
the function
\ben
U(P,Q,T) = u(p,q,t)
\label{1-18}
\een
satisfies
\ben
[\partial_T+ P\partial_Q \mp  Q\partial_P]U(P,Q,T) = 0
\label{1-19}
\een
provided (\ref{1-12}) holds.

This transformation is in fact a canonical transformation. Therefore the condition (\ref{1-15a}) that $u(p,q,t)$ describes $N$ fermions transforms into the condition 
\ben
\int dPdQ~u(P,Q,T) = N
\een
The equation (\ref{1-19}) is the equation satisfied by the phase space density operator for a system of fermions which is in an external harmonic (or inverted harmonic) potential with unit mass and unit frequency. Therefore once we know the solution for this latter case, we can find a solution of the time dependent case in terms of a solution of the equation (\ref{1-11}).

\section{Quantization and the "in" state}
\label{sec-three}

The quantization of the fermionic field theory proceeds in a standard fashion. Given a complete set of modes $\{ \psi_n (x,\tau) \}$ which solve the equations of motion the Heisenberg picture field operators may be expressed as
\bea
\psi (x,\tau) & = & \sum_{n=0}^\infty a_n~\psi_n (x,\tau) \nn \\
\psi^\dagger (x,\tau) & = & \sum_{n=0}^\infty a^\dagger_n~\psi^*_n (x,\tau)
\label{2-1}
\eea
where the oscillators satisfy the standard anti-commutation relations 
\ben
\{ a_m , a^\dagger_n \} = \delta_{mn}~~~\{ a_n, a_m \} = \{ a^\dagger_m, a^\dagger_n \} = 0 
\label{2-2}
\een
Different choices of modes determine different inequivalent quantizations related by Bogoliubov transformations. 

We will be interested in profiles of $m(t),\nu(t)$ such that they approach constant values $m_{in}$ and $\nu_{in}$ as $t \rightarrow -\infty$, and their time derivatives approach zero. Furthermore we will have choices of $m(t)$ such that when $t \rightarrow -\infty$, one also has $\tau \rightarrow -\infty$. In particular our choices of $m(t)$ are such that as $t \rightarrow -\infty$, we have $m(t) \rightarrow m_{in}$ so that $\tau \rightarrow \frac{1}{m_{in}} t$.  The equation (\ref{1-11}) means that $\rho(t) = \rho(\tau)$ has the initial condition
\ben
{\rm Lim}_{\tau \rightarrow -\infty}\rho (\tau) = \rho_{in} = \frac{1}{\sqrt{m_{in}\nu_{in}}} = \frac{1}{\sqrt{\omega_{in}}}
\label{1-14-3}
\een
The corresponding solution $\psi_n$ in (\ref{2-1}) must then have the property that this is positive frequency in the far past,
\ben
{\rm Lim}_{\tau \rightarrow \infty} \psi_n (x,\tau) \sim e^{-i\alpha \tau}
~~~~~\alpha > 0
\een
We will consider the Heisenberg picture state which is the "in" ground state,
\bea
a_n |in\rangle &= & 0 ~~~~~~~~~n \geq N \nn \\
a^\dagger_n |in\rangle & = & 0~~~~~~~~~0 \leq n \leq N-1
\label{2-3}
\eea

\subsection{Observables}

The observables we will be interested in are the expectation value of the quenched operator and the entanglement entropy. We will now show that both these quantities can be expressed in terms of the corresponding quantities in the time independent problem and the function $\rho(\tau)$.

In the following we will consider the expectation value of the operator
\ben
\cO (\tau) = \int_{-\infty}^\infty dx~x^2 \psi^\dagger (x,\tau) \psi (x,\tau) 
\label{2-4}
\een
This is the operator which comes multiplied by the time dependent coupling $\omega^2(\tau)$ once the theory is expressed in the time variable $\tau$. In the spirit of response theory, the expectation value then measures the response of the system to the external driving. $\langle \cO (\tau) \rangle$  of our problem can be expressed simply in terms of the expectation value of the quenched operator in an auxiliary problem of a harmonic oscillator with unit mass and frequency, using (\ref{1-14-2})
\ben
\langle in |\cO (\tau)| in\rangle = \sum_{n=0}^{N-1}\int_{-\infty}^\infty dx~x^2~\psi^*_n(x,\tau) \psi_n(x,\tau) 
\een
Using (\ref{1-14-1}) and (\ref{1-14-2}) this becomes, after a change of variables,
\ben
\langle in|\cO(\tau)|in \rangle = \rho^2(\tau) \sum_{n=0}^{N-1}\int_{-\infty}^\infty dY Y^2 \phi^*_n(Y) \phi_n(Y)= \rho (\tau)^2 \sum_{n=0}^{N-1} (n+1/2) = \frac{N^2}{2} \rho(\tau)^2
\label{2-6}
\een
where we have used the fact that the integral on the right hand side is the expectation value of the potential energy of a single harmonic oscillator with unit frequency in the state with quantum number $n$, and used the standard result.

For fermionic systems, the entanglement entropy of a subregion $A$ has an expansion in terms of cumulants of the particle number distribution \cite{cumulants,peschel,satya}. In the leading order of large $N$ the dominant term is the variance of the expectation value of the particle number in $A$,
\ben
S_A (\tau)= \frac{\pi^2}{3} [ \langle N_A(\tau)^2\rangle - \langle N_A(\tau)\rangle^2 ]
\label{2-7}
\een
where the operator $N_A$ is given by
\ben
N_A (\tau)  = \int_A dx \psi^\dagger (x,\tau) \psi (x,\tau)
\label{2-8}
\een
where the integral is over the region $A$.

This simplifies for the "in" state. Using the mode expansion (\ref{2-1}) and the state defined in (\ref{2-3}) it may be easily shown that
\ben
S_A (\tau) = \langle in|N_A(\tau)|in\rangle - \int_A dx \int_A dy |C(x,y,\tau)|^2
\label{2-9}
\een
where
\ben
C(x,y,\tau) = \langle in|\psi^\dagger (x,\tau) \psi (y,\tau)|in\rangle
\label{2-10}
\een
This quantity can be also expressed entirely in terms of the expectation value of the phase density operator as follows
\bea
S_A & = &{1\over2\pi} \int_{-\infty}^\infty dp  \int_A dx \, \langle in|u(p,x,\tau)|in\rangle - \nn \\
& & {1\over(2\pi)^2}\int_{-\infty}^\infty dp_1 dp_2 \int_A dx dy~ e^{-i(p_2-p_1)(x-y)}~\langle in|u(p_1,(x+y)/2,\tau)|in\rangle \langle in|u(p_2,(x+y)/2,\tau)|in\rangle\nn\\
\label{2-11}
\eea
Expressing the above expectation values in terms of the mode functions one has
\bea
\langle in|N_A(\tau)|in\rangle & = & \int_A dx~\sum_{n=0}^{N-1} \psi_n^* (x,\tau) \psi_n (x,\tau) \nn \\
C(x,y,\tau) & = & \sum_{n=0}^{N-1} \psi_n^* (x,\tau) \psi_n (y,\tau)
\label{2-12}
\eea
Using (\ref{1-14-2}) it then follows that the entanglement entropy can be expressed in terms of the entanglement entropy of a rescaled region in the ground state of the theory with a constant mass and frequency.
If the subregion $A$ is defined by $a \leq x \leq b$ then the rescaled subegion is defined by 
\ben
S_A[\omega(\tau)] = S_{A_P}[\omega=1]~~~~~~~~~
A_P : \frac{a}{\rho(\tau)} \leq x \leq \frac{b}{\rho(\tau)}
\label{2-13}
\een

\section{Results for fermions in Harmonic Oscillator Potential}
\label{sec-five}

For the right side up harmonic oscillator, the two independent solutions of the equation (\ref{1-13}) may be therefore chosen to be such that
\ben
{\rm Lim}_{\tau \rightarrow -\infty} f(\tau) = \frac{1}{\sqrt {2\omega_{in}}}e^{-i\omega_{in}\tau}~~~~~~g(\tau) = [f(\tau)]^*
\label{1-14-4}
\een
To ensure that $\rho(\tau)$ is real we then need to choose
\ben
A = C = 0~~~~~~~~~B = 1
\label{1-14-5}
\een
Therefore for this solution we have
\ben
\rho (\tau) = \sqrt{2}~|f(\tau)|
\label{1-14-6}
\een
This yields the final form of the solution 
\begin{equation}
\begin{split}
\psi_n (x,\tau) = \frac{1}{\sqrt{2^n n !}} \left[ \frac{1}{\pi \rho (\tau)^2} \right]^{1/4}~{\rm exp} \left[ -i(n+1/2) \int^\tau \frac{dt'}{\rho(\tau)^2}  \right]~ \hfill \\
{\rm exp}\left[\frac{i}{2} \left( \partial_\tau \log \rho(\tau) + \frac{i}{\rho(\tau)^2} \right)x^2 \right]~H_n (x/\rho(\tau))
\end{split}
\label{1-14-7}
\end{equation}
where $H_n(x)$ denotes the n-th order Hermite polynomial. This solution approaches the normalized solutions of the Schrodinger equation with a frequency $\omega_{in}$ as $\tau \rightarrow -\infty$. The oscillators in (\ref{2-1}) with these modes are in the "in" oscillators.

We now provide exactly solvable quench protocols for fermions with a fixed mass $m$ in a harmonic oscillator potential with time dependent frequencies. The two times $t$ and $\tau$ are then related by $\tau = t/m$.

\subsection{Cis-Critical Protocol}
The first protocol is a cis-critical-protocol (CCP). As described in the introduction in such a protocol the system starts from a gapped phase, approaches a critical point and then turns back to another constant value. In this work we choose a protocol where the initial and the final values are the same. More specifically we choose
\ben
\omega(\tau)^2 = \omega_0^2~\tanh^2 (\tau/\dt)
\label{3-1}
\een
This corresponds to a trap which is smoothly removed for a finite interval of time and then re-introduced.

The solution to the equation (\ref{1-13}) which behaves as $e^{-i\omega_0 \tau}$ is then given by
\ben
\begin{split}
f_{CCP}(\tau) = \frac{1}{\sqrt{2\omega_0}} & \frac{2^{i\omega_0\delta t}\text{cosh}^{2\alpha}(\tau/\delta t)}{E_{1/2}\tilde{E}'_{3/2}-E'_{1/2}\tilde{E}_{3/2}}   \times \hfill \\
& \left\{ \tilde{E}'_{3/2}~ {_2F_1}(a,b;\frac{1}{2}; -\text{sinh}^2 \frac{\tau}{\delta t}) + E'_{1/2} \text{sinh}\frac{\tau}{\delta t}~ {_2F_1}(a+\frac{1}{2},b+\frac{1}{2};\frac{3}{2};-\text{sinh}^2\frac{\tau}{\delta t})   \right\}
\end{split}
\label{3-2}
\een
where we defined
\begin{equation}
\begin{split}
\alpha = \frac{1}{4}[1+\sqrt{1-4\omega_0^2\delta t^2}]  \hfill \\
a = \alpha - \frac{i}{2} \omega_0 \delta t, 
b= \alpha + \frac{i}{2} \omega_0 \delta t  \hfill \\
E_{1/2} = \frac{\Gamma(1/2)\Gamma(b-a)}{\Gamma(b)\Gamma(1/2-a)}, 
\tilde{E}_{3/2} = \frac{\Gamma(3/2)\Gamma(b-a)}{\Gamma(b+1/2)\Gamma(1-a)} \hfill \\
E_c'= E_c(a \leftrightarrow b)
\end{split}
\label{3-3}
\end{equation}
The key function $\rho(\tau)$ is then given by (\ref{1-14-6}) with $f(\tau)$ given by (\ref{3-2}).

\subsection{End Critical Protocol (ECP)}

Another solvable quench protocol is the end critical protocol where the initial theory is a harmonic oscillator with a frequency $\omega_0$ which monotonically descreases smoothly to a vanishing frequency at infinitely late times. This corresponds to a {\em smooth} release from a harmonic trap. 

\begin{figure}
\centering
\subfloat[$\omega_0\delta t=1$(CCP)]{
\includegraphics[width=0.45\textwidth]{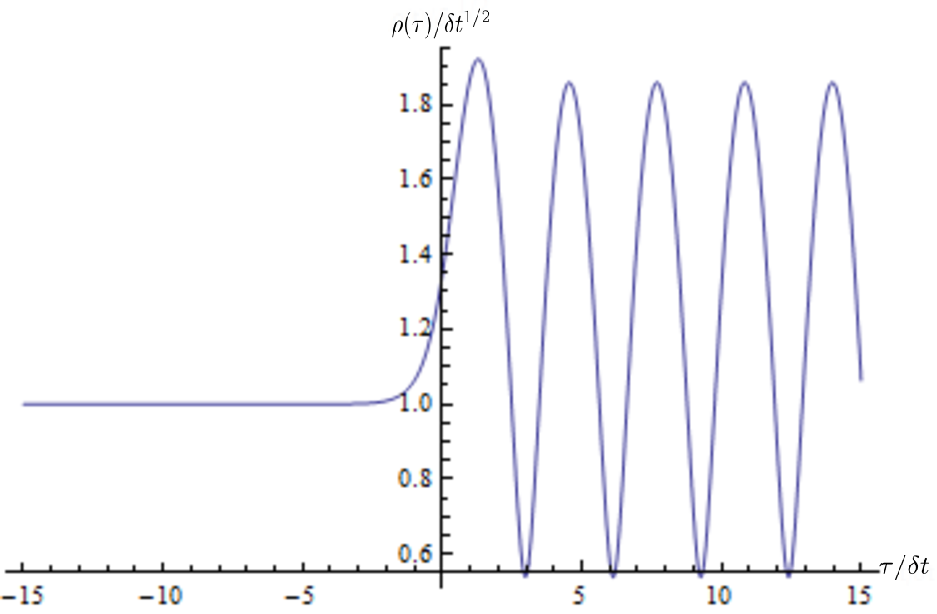}}
\subfloat[$\omega_0\delta t=1$(ECP)]{
\includegraphics[width=0.45\textwidth]{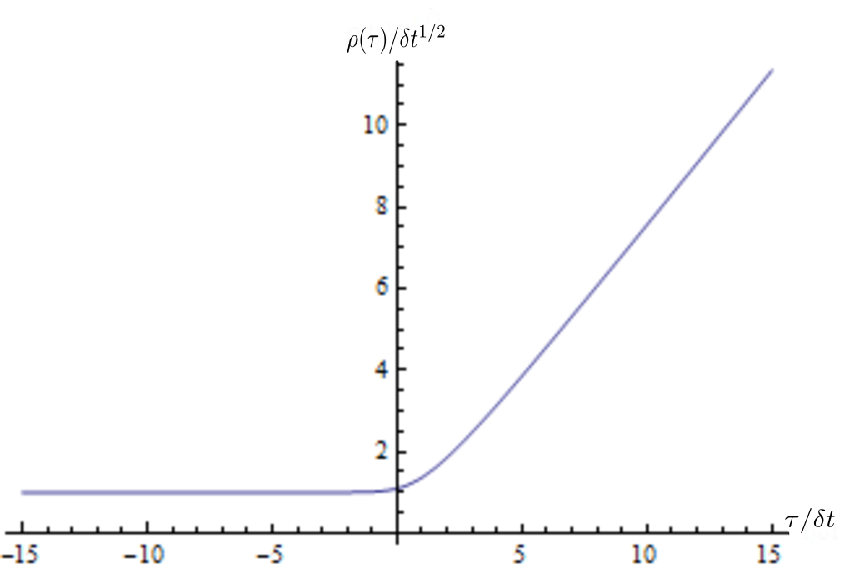}}\\
\subfloat[$\omega_0\delta t=100$(CCP)]{
\includegraphics[width=0.45\textwidth]{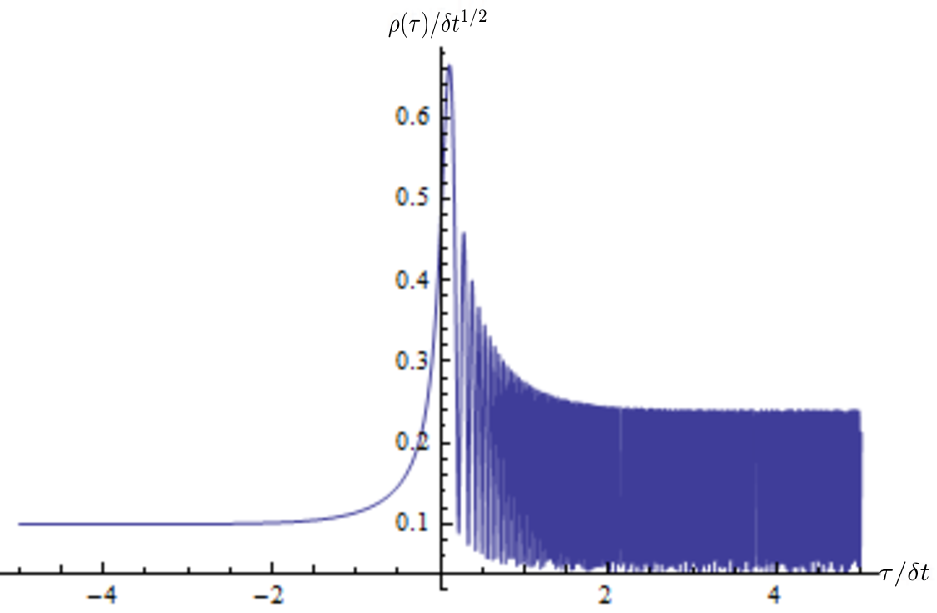}}
\subfloat[$\omega_0\delta t=100$(ECP)]{
\includegraphics[width=0.45\textwidth]{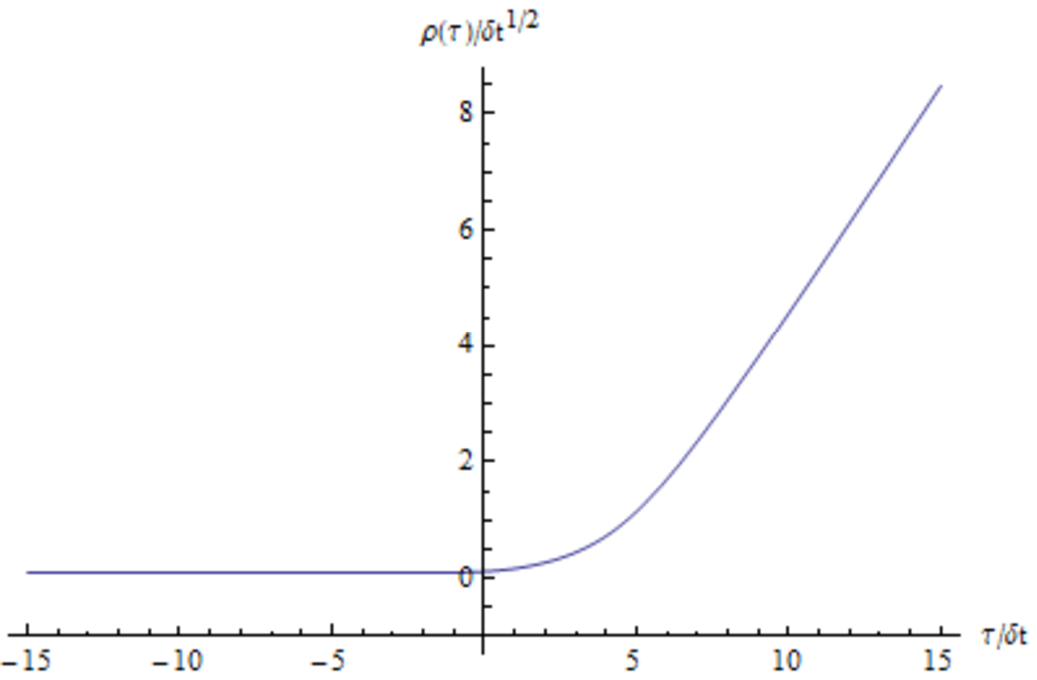}}\\
\subfloat[$\omega_0\delta t=0.01$(CCP)]{
\includegraphics[width=0.45\textwidth]{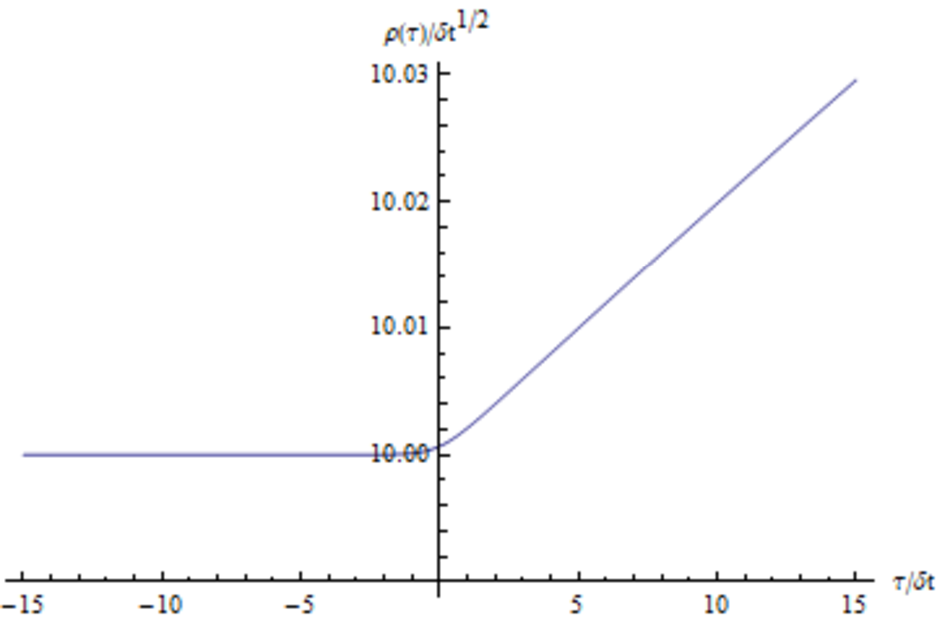}}
\subfloat[$\omega_0\delta t=0.01$(ECP)]{
\includegraphics[width=0.45\textwidth]{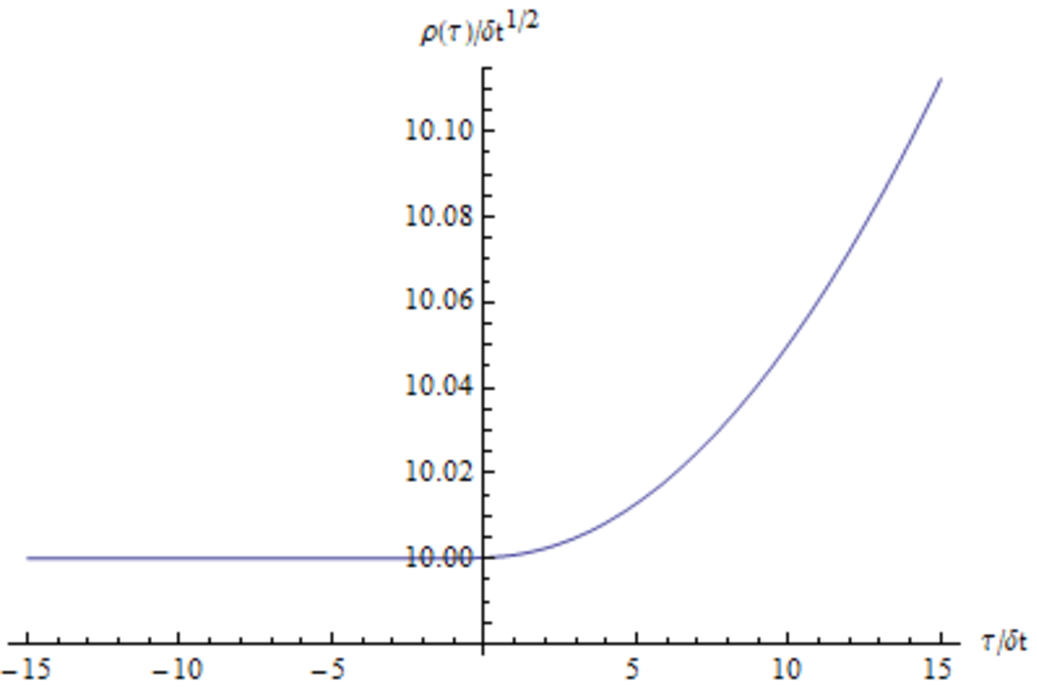}}\\
\caption[Relation between $\rho(\tau)$ and $\tau$]{relation between $\rho(\tau)$ and $\tau$ in various $\omega_0 \delta t$ cases.}
\label{fig-two}
\end{figure}

Consider the slightly more general protocol
\begin{equation}
\omega^2(\tau)=\omega_0^2 \left(a+b \tanh \frac{\tau}{\delta t} \right)
\label{4-1}
\end{equation}
with the real constants $a,b$ chosen such that $a > b$ to ensure reality of $\omega (\tau)$.
Then the "in" solution of the equation (\ref{1-13}) is given by
\begin{equation}
\begin{split}
f_{ECP}= \frac{1}{\sqrt{2\omega_{in}}} & \text{exp}[-i\omega_+ \tau-i\omega_- \delta t \text{log}(2 \text{cosh}(\tau/\delta t))] \hfill \\
& _2F_1[1+i\omega_-\delta t, i\omega_-\delta t ; 1-i\omega_{in}\delta t; \frac{1}{2}(1+\text{tanh}(\tau/\delta t ))]  \hfill \\
\end{split}
\label{4-2}
\end{equation}
where we defined
\begin{equation}
\begin{split}
\omega_{in}=\omega_0 \sqrt{a-b}, \hfill \\
\omega_{out}= \omega_0 \sqrt{a+b}, \hfill \\
\omega_{\pm}=\frac{1}{2}(\omega_{out} \pm \omega_{in})
\end{split}
\end{equation}
The end critical protocol we consider has $a = -b = \frac{1}{2}$.
The function $\rho(\tau)$ which determines the time dependence of the observables considered above is shown in Figure \ref{fig-two} for both these types of protocol. 

At early times $\rho (-\infty) = \frac{1}{\sqrt{2\omega_{in}}}$. For ECP $\rho(\tau)$ monotonically increases and behaves as $\rho(\tau) \sim \tau$ at large $\tau$. For CCP $\rho(\tau)$ initially increases and then starts oscillating. At late times these oscillations are around a mean value which is roughly the initial value 
$\frac{1}{\sqrt{2\omega}}$ with an amplitude which remains constant in time and with a frequency approximately given by $\omega_0$.

\section{The response and scaling : CCP}
\label{sec-six}

In this section we present the results of the expectation value of the quenched operator $\cO = \int dx x^2 \psi^\dagger \psi$ at early times for CCP (equation (\ref{3-1})) and investigate their scaling behavior in various regimes. The details of the analytic approximations which lead to these results are given in Appendix \ref{appA}.

\subsection{Slow Quench Regime}

In the slow quench regime $\omega_0 \dt \gg 1$ we can use the asymptotic form of gamma functions
\ben
\Gamma(z) \sim \sqrt{2\pi}e^{-z+(z-\frac{1}{2})\log z}, z \to \infty
\end{equation}
to obtain $\rho (\tau=0)$. The leading expression for the one point function $\langle \cO\rangle$ at $\tau = 0$ is, using (\ref{2-6}),
\ben
\langle \cO (0) \rangle \sim \frac{\sqrt{\pi}}{2}N^2\sqrt{\frac{\delta t}{\omega_0}}
\label{3-4}
\end{equation}
This result is consistent with Kibble-Zurek scaling. The Landau criterion with the instantaneous frequency given by (\ref{3-1}) leads to 
\ben
 \frac{1}{\omega_0 \delta t} {\rm cosech}^2 (\tau_{KZ}/\dt) = 1
\label{3-5}
\een
which defines the Kibble-Zurek time $\tau_{KZ}$.
We expect a scaling behavior only when $\tau_{KZ} \ll \dt$. In this regime (\ref{3-5}) leads to
\ben
\tau_{\text {KZ}}= \sqrt{\frac{\delta t}{\omega_0}}
\label{3-6}
\een
The condition $\tau_{KZ} \ll \dt$ then becomes consistent with the slow quench condition $\omega_0 \dt \gg 1$. This leads to the instantaneous frequency at the Kibble-Zurek time, 
\ben
 \omega_{KZ}^2 =  \frac{\omega_0}{\dt}
\label{3-7}
\een
According to the Kibble-Zurek argument $\rho(\tau)$ in the middle of the quench (which is $\tau=0$) is roughly equal to its value at $\tau = \tau_{KZ}$. Since the system is approximately adiabatic at $\tau = \tau_{KZ}$ this is in turn roughly equal to $\rho_{adia}(\tau_{KZ})$, the value of $\rho$ for the fermions in a harmonic oscillator potential with a constant frequency $\omega_{KZ}$. From (\ref{1-11}) this is simply 
\ben
\rho(\tau_{KZ}) \sim \rho_{adia}(\tau_{KZ}) = \frac{1}{\sqrt{\omega_{KZ}}}
\een
leading to
\ben
\langle\cO\rangle \sim \frac{N^2}{2} \sqrt{\frac{\delta t}{\omega_0}}
\label{3-8}
\een
which is in agreement with the result from the exact solution (\ref{3-4}) upto a numerical factor.

\subsection{Fast Quench Regime}

We now consider the regime $\omega_0 \dt \ll 1$. While we have the exact answer anyway, we are able to approximate the answer by suitable expansions and obtain analytic expressions when we have in addition $\omega_0 \tau \ll 1$ . The latter are useful to make a comparison with perturbation calculations.

First consider the response at a time $\tau$ which is in the range
\ben
\omega_0 \tau \ll \omega_0 \dt \ll 1
\label{3-9}
\een
In this case, for the CCP (equation (\ref{3-1}) we get an expression (see Appendix \ref{A-1}, equations (\ref{a8})-(\ref{a18}),
\ben
\langle \cO (\tau)\rangle \approx \frac{N^2}{2\omega_0}\left\{  1+2\text{log}2 \cdot \omega_0^2\delta t^2 
+2\omega_0^2 \delta t \cdot \tau  +\mathcal{O}(\omega_0^4\delta t^4, \frac{\tau^2}{\delta t^2})\right\} 
\label{3-10}
\end{equation}
This is the response at early times. At late times, (see equations (\ref{a19}) to (\ref{a28}))
\ben
\omega_0 \dt \ll \omega_0 \tau \ll 1
\label{3-11}
\een
one gets instead
\ben
\langle \cO \rangle \sim \frac{N^2}{2\omega_0} \left( 1 +2\omega_0\delta t \text{sin} 2\omega_0 \tau  +\mathcal{O}(\omega_0^2 \delta t^2) \right) 
\label{3-12}
\end{equation}

These results should also follow from usual time dependent perturbation theory. Let us discuss this for a general perturbation $\ \delta \omega(\tau)^2$ from the initial value. 
The leading term in the perturbation expansion is
\ben
\langle \cO (\tau) \rangle = \langle\cO (-\infty)\rangle +\frac{1}{2} \int_{-\infty}^{\tau} d\tau' \int dx \int dx' ~(xx')^2 \delta \omega(\tau')^2 \langle 0|[ \psi^\dagger (x,\tau)\psi(x,\tau) , \psi^\dagger (x',\tau')\psi(x',\tau') ]|0\rangle_{\omega_0}
\label{3-13}
\een
where $\langle \rangle_{\omega_0}$ denotes the expectation value in the ground state of the theory at $\tau \rightarrow -\infty$ which is the harmonic oscillator with a constant frequency $\omega_0$ and 
\ben
\delta \omega(\tau)^2 = \omega(\tau)^2 - \omega_0^2
\label{3-14a}
\een
The Green's function which appears in the linear response can be calculated. The result is
\bea
G(\tau,\tau') & = & \theta (\tau-\tau')\int dx\int dx'~x^2~(x')^2~\langle 0|[ \psi^\dagger (x,\tau)\psi(x,\tau) , \psi^\dagger (x',\tau')\psi(x',\tau') ]|0\rangle_{\omega_0} \nonumber \\
& = & - \theta(\tau-\tau') \frac{N^2}{\omega_0^2}\sin[2\omega_0(\tau-\tau')] 
\label{3-14}
\eea
Now consider evaluating the response at a time which is of the order of $\tau \sim +\dt$. In the fast quench regime $\omega_0\dt \ll 1$. The limits of the integral in (\ref{3-14}) can be replaced by $(-\dt,\dt)$. Suppose the form of $\delta \omega(\tau)^2$ is 
$\delta \omega(\tau)^2 = \delta \omega_0^2 f (\tau/\dt)$ where $\dt$ is the time scale of the quench and $f(x)$ some smooth function. Using the above form of the Green's function the linear response becomes in the fast quench regime
\bea
\langle\cO (\tau) \rangle - \langle\cO (-\infty)\rangle & \sim & \delta \omega_0^2 \frac{N^2}{2\omega_0^2}\int_{-\dt}^{\dt} d\tau'~f(\tau'/\dt) \sin[2\omega_0 (\tau-\tau')] \nn \\
& \sim & \delta \omega_0^2 \frac{N^2}{2\omega_0^2}\omega_0 \dt^2
\label{3-15}
\eea
In the protocol we are using $\delta \omega_0^2 = \omega_0^2$ and $f(\tau/\dt) = {\rm sech}^2(\tau/dt)$. We therefore reproduce the scaling in (\ref{3-10}).

\subsection{The exact response} 
\begin{figure}
\centering
\setlength{\abovecaptionskip}{0 pt}
\centering
\includegraphics[scale=0.6]{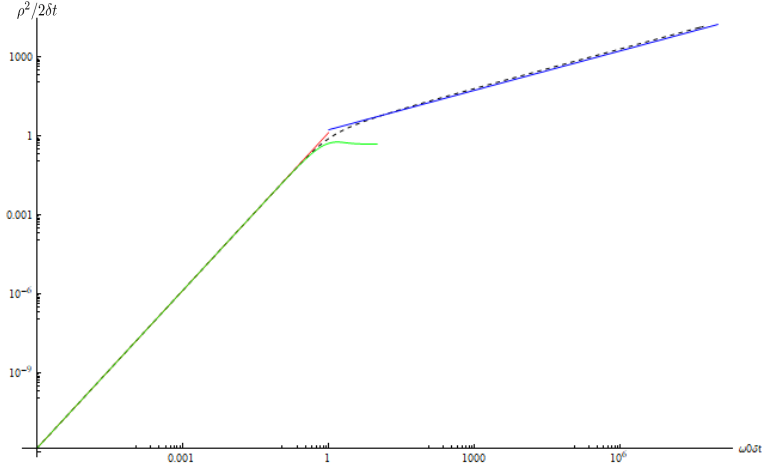}
\caption{(Colour online) The response $\rho^2/(2\delta t) = \langle \cO \rangle/(N^2 \delta t)$ as a function of $\omega_0 \delta t$ for CCP when $\tau =0$.  
The black dashed curve is the exact result obtained by using (\ref{3-2}). The blue curve is the leading Kibble Zurek result for $\omega_0 \delta t \gg 1$, i.e. Eq. (\ref{3-8}).
The red curve is the leading behavior when $\omega_0 \delta t \ll 1$, i.e. Eq. (\ref{3-10}).
The green curve is the perturbation expansion result i.e. Eq. (\ref{3-15}).}
 \label{fig-one}
\end{figure}

The exact response for CCP is shown in Figure \ref{fig-one}.
This shows that the analytic approximatations in the fast and slow regime agree very well with the exact answer, and the transition between the two regimes is rather sharp.

\section{The response and scaling : ECP}

The investigation of the scaling behavior for the ECP case (\ref{4-1}) follows along lines similar to CCP.

\subsection{Slow Quench Regime}

In the slow quench regime  ($\omega_0\dt \gg 1$) one expects a Kibble Zurek scaling. 
For the protocol (\ref{4-1}) the Landau criterion determining the Kibble-Zurek time $t_{KZ}$ becomes
\ben
\frac{1}{\cosh^2 (\tau_{KZ}/\dt)(1-\tanh (\tau_{KZ}/\dt))^{3/2}}\sim \omega_0\dt
\label{5-1}
\een
For $\omega_0\dt \gg 1$ the solution can appear only at late times. This yields
\ben
\tau_{KZ} \sim \dt \log(\omega_0\dt)
\label{5-2}
\een
In this case the instantaneous gap vanishes in the infinite future. This means that in the slow quench regime adiabaticity will fail at late times. The frequency at this time is
\ben
\omega_{KZ} = \omega_0 \sqrt{\frac{1-\tanh (\tau_{KZ}/\dt)}{2}} \sim \frac{1}{\dt}
\label{5-3}
\een
Therefore the standard Kibble Zurek argument would predict that the response at late times is given by
\ben
\langle \cO \rangle \sim \frac{1}{2}N^2 \rho(\tau_{KZ})^2 =
\frac{N^2}{2\omega_{KZ}} \sim \frac{1}{2}N^2 \dt
\label{5-4}
\een
At a time earlier than the 
Kibble Zurek time i.e. when
\ben
\tau < \dt \log(\omega_0\dt),
\een
the adiabatic approximation is valid. Therefore if one measures the response at some fixed value of $\tau/\dt = \zeta$ we should have 
\ben
\langle \cO \rangle \sim  \frac{N^2}{2\omega_0 \sqrt{\frac{1-\tanh(\tau/\dt)}{2}}} \sim  \frac{N^2}{\sqrt{2}\omega_0 \sqrt{1-\tanh \zeta}}
\label{5-5}
\een

This expectation needs refinement. Using the exact solution we can perform an expansion for $\omega_0\dt \gg 1$ and for $\tau \gg \delta t \log \omega_0 \delta t  $.
We find that the leading term of $\rho^2(\tau)$ is 
\begin{equation}
\rho^2(\tau) \sim \delta t \left[ \frac{2}{\pi} \left(- \log \omega_0 \delta t + \log 2-\gamma_E +\frac{\tau}{\delta t} \right)^2  + \frac{\pi}{2}  \right] \sim \cO(1)
\label{a-9}
\end{equation}
The additional logarithmic dependence is not easily visible from the naive Kibble-Zurek argument.

\subsection{Fast Quench Regime}

In the fast quench regime one can get an analytic expression 
\begin{equation}
\begin{split}
\rho^2(\tau) \sim &   \omega_0 \delta t^2 \left( -\frac{\zeta(3)}{4}\omega_0^2\delta t^2 +\frac{\tau}{\delta t} \right)^2 +    \frac{1}{\omega_0 } = \frac{1}{\omega_0} + \omega_0 \tau^2 - \frac{\zeta(3)}{2} (\omega_0\dt)^3 \tau.\hfill \\
\end{split}
\label{a-8}
\end{equation}
at late times, i.e. $\omega_0 \delta t \ll \omega_0 \tau \ll 1$.
Details of calculation which leads to (\ref{a-8}) are summarized in Appendix \ref{A-2}.

The limit $\dt \rightarrow 0$ is smooth. In this limit the expression (\ref{a-8}) reduces to the result which is obtained in an abrupt quench where the frequency suddenly changes from $\omega_0$ to zero,
\ben
\rho^2_{abrupt}(\tau) = \frac{1}{\omega_0} [ 1 + (\omega_0\tau)^2]
\label{a-8a}
\een
In relativistic theories this limit is non-trivial because of UV divergences, as discussed in \cite{dgm1}.

Once again the answer should be obtainable by a perturbation expansion in $\omega_0\dt$.
Again let $\delta \omega(\tau)^2 = \omega_0^2 f(\tau/\delta t)$, where 
\begin{equation}
f(x) = \left\{
{\begin{array}{*{20}{c}}
 0, & x <-1; \\
 \frac{1+x}{2}, & -1 \le x \le 1;\\
 1 , & x >1. \\
\end{array} } \right.
\end{equation}
Then at late times,
\bea
\langle\cO (\tau) \rangle - \langle\cO (-\infty)\rangle & \sim &  \omega_0^2 \frac{N^2}{2\omega_0^2}\int_{-\dt}^{\tau} d\tau'~f(\tau'/\dt) \sin[2\omega_0 (\tau-\tau')] \nn \\
& \sim &   \frac{N^2}{2\omega_0}\sin^2 \omega_0 \tau.
\eea
Thus $\omega_0 \tau \ll 1$ the perturbation expansion gives a good approximation.

\subsection{The exact response} The above discussion shows that for the ECP it is useful to look at the response for a {\em fixed} value of $\tau/\dt = \zeta$. Our analytic approximations then predict
\[ \frac{2\langle \cO \rangle}{N^2 \dt} = \left\{
\begin{array}{lr}
\frac{1}{\omega_0\dt}+ (\omega_0 \dt) \zeta^2 & : \omega_0\dt \ll 1 \nn \\
{\rm constant} & : 1 \ll \omega_0 \dt \ll e^\zeta \nn \\
 \frac{\sqrt{2}}{(\omega_0 \dt)\sqrt{1-\tanh(\zeta)}} & : \omega_0\dt \gg e^{\zeta} \\
\end{array}
\right.
\label{5-55a}
\]

\begin{figure}
\centering
\setlength{\abovecaptionskip}{0 pt}
\centering
\includegraphics[scale=0.8]{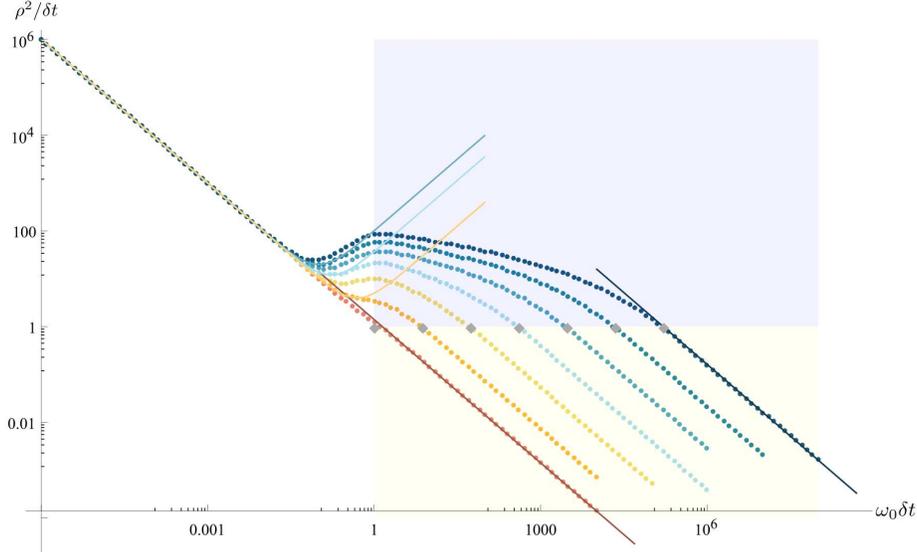}
\caption{(Colour online) The response $\rho^2(\tau)/\dt$ as a function of $\omega_0 \delta t$ for ECP. 
The dots are the exact results obtained by using (\ref{4-2}) for fixed values of $\zeta = \tau/\dt = 0,2,4,6,8,10,12$ which are colored from red to blue respectively. The grey dot on each curve corresponds to $\omega_0\dt = e^{\zeta}$ for that particular $\zeta$. Thus all points in the yellow shaded region are in the adiabatic regime. The points which lie in the blue shaded region have $1 < \omega_0\dt <e^\zeta$. For  larger values of $\zeta$ there is a small window in this regime where $\rho^2(\tau)/\dt$ is roughly constant  which is the expectation from Kibble Zurek scaling. The slight increase is consistent with the logarithmic term in (\ref{a-9}).
The dark red and dark blue solid lines are the linear fitting ($\log y= P \log x +Q$) results of red ($\tau/ \delta t =0$) and blue dots ($\tau/ \delta t =12$) when $\omega_0 \delta t \gg e^{\tau/\delta t}$ (yellow region), respectively. Both the slopes $P$ are approximately $-1$.
The orange, blizzard blue and light blue solid curves in the fast quench regime ($\omega_0 \delta t \ll 1$) are the sudden quench result (\ref{a-8a}) for $\tau/ \delta t=2,6,10$, respectively. For $\omega_0\dt < 1$ the data points lie on these solid lines. For $\omega_0\dt >1$ they continue to lie on the solid lines for a while and then depart from them, reflecting the $O(\omega_0^3\dt^3)$ terms in (\ref{a-8}).}
 \label{fig-four}
\end{figure}

\begin{figure}
\centering
\setlength{\abovecaptionskip}{0 pt}
\centering
\includegraphics[scale=0.8]{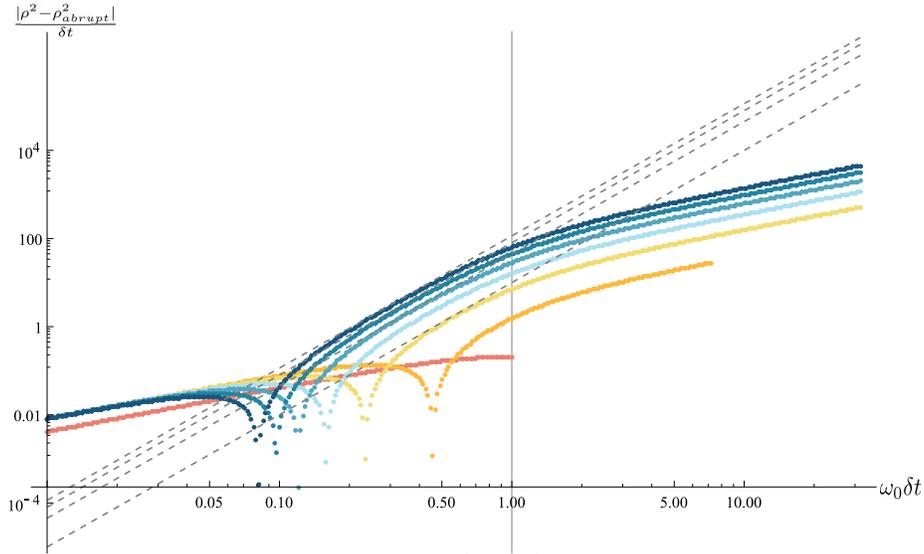}
\caption{(Colour online) The response $\frac{|\rho^2(\tau)-\rho^2_{abrupt}(\tau)|}{\dt}$ as a function of $\omega_0 \delta t$ for ECP. 
The dots are the exact results obtained by using (\ref{4-2}) for fixed values of $\zeta = \tau/\dt = 0,2,4,6,8,10,12$ which are colored from red to blue respectively. The vertical gridline $\omega_0 \delta t=1$ is the threshold between fast quench and slow quench. The dashed lines are a set of cubic functions $y=ax^3$, where $a=10,45,80,115$ from the lowest one to the highest one, respectively to compare with the leading term in (\ref{a-8}).}
 \label{fig-five}
\end{figure}

Figure \ref{fig-four} shows how the exact result compares with the above expectations. Here we plot the quantity $\rho^2 /\dt =2 \langle \cO \rangle/(N^2 \dt)$ as a function of $\omega_0\dt$ for different values of $\zeta$. For very small $\omega_0\dt$ one reproduces the abrupt quench result. For slightly larger $\omega_0\dt$ we can see the fast quench correction predicted in (\ref{a-8}). 
To investigate the behavior in the fast quench regime, it is useful to subtract the abrupt quench response. The quantity $|\rho^2(\tau) - \rho^2_{abrupt}(\tau)|/\dt$ is plotted in Figure \ref{fig-five}. This quantity is close to zero (and slightly negative) for sufficiently small $\omega_0\dt$. For larger $\omega_0\dt$ this becomes positive and in a reasonable range of $\omega_0 \dt$ this is consistent with the $(\omega_0 \dt)^3$ term in the fast quench response, equation (\ref{a-8}) which are shown by solid lines. Note that the cusps in the data appear because the quantity $\rho^2(\tau) - \rho^2_{abrupt}(\tau)$ changes sign and we are plotting the absolute value - there is nothing singular here.

For sufficiently large values of $\omega_0\dt$ this quantity is proportional to $1/(\omega_0\dt)$ with a proportionality constant which depends on $\zeta$, as expected from an adiabatic response. There is a small window in the intermediate regime where $\rho^2(\tau)/\dt$ is roughly constant upto a logarithmic dependence as in (\ref{a-9}).

\clearpage

\section{Entanglement Entropy}
\label{sec-eight}

In this section we present the results for the entanglement entropy of a subregion, its scaling at early times and the time dependence at late times. As argued above, the entanglement entropy in a given subregion for a time dependent frequency can be expressed entirely in terms of the entanglement entropy of a {\em scaled} subregion for the system at fixed unit frequency, with the scaling factor given by $\rho(\tau)$ (eqn (\ref{2-13})). In the following we will examine the behavior of the entanglement entropy for a subregion $-a \leq x \leq a$. We will also be interested in the limit $N \gg 1$ so that we can use the expression (\ref{2-9}). 

We will be interested in the entanglement entropy for a subregion size
\ben
\frac{a}{\rho(\tau)} \ll \sqrt{N}
\label{5-7}
\een
For ECP the function $\rho(\tau)$ monotonically increases with time, so this condition is equivalent to the condition $\sqrt{\omega_0} a \ll 1$ since $\rho(-\infty) = \frac{1}{\sqrt{\omega_0}}$ - the monotonicity then implies that once we impose (\ref{5-7}) at the initial time, this will continue to hold for all times. 
For CCP the function $\rho(\tau)$ oscillates roughly around $\rho (-\infty)$ with an amplitude which is roughly constant in time : once we pick a value of $a$ such that this condition is satisfied at some sufficiently large time, this will continue to be satisfied for all times. 

The expression for entanglement entropy at large $N$ can be written down using (\ref{2-9}, \ref{2-10}) and (\ref{2-13}) by using the Christoffel-Darboux formula for orthogonal polynomials
\begin{equation}
 \sum _{k=0}^{n}{\frac {H_{k}(x)H_{k}(y)}{k!2^{k}}}={\frac {1}{n!2^{n+1}}}\,{\frac {H_{n}(y)H_{n+1}(x)-H_{n}(x)H_{n+1}(y)}{x-y}}.
\end{equation}
This leads to
\bea
\langle N_A \rangle & = & \frac{1}{\Gamma(N) 2^N \sqrt{\pi}} \int_{A_P} d\xi e^{-\xi^2}~[H_{N-1}(\xi)H_N^\prime(\xi) - H_{N-1}^\prime(\xi) H_N(\xi)]  \\
\int_{A_P \times A_P} {\text d} x {\text d} y |C(x,y) |^2  &=& - \frac{1}{\pi 2^{2N}(\Gamma(N))^2}
\int_{A_P \times A_P} \ d\xi d\eta e^{-(\xi^2+\eta^2)} \left(\frac {H_{N-1}(\eta)H_{N}(\xi)-H_{N-1}(\xi)H_{N}(\eta)}{\xi-\eta} \right)^2 \nn \\
\label{cxy_ex}
\eea
where the notation $A_P \times A_P$ means that the integrals go over the range defined by $A_P$.
These expressions simplify in two regimes. First consider the regime
\ben
\frac{1}{\sqrt{N}} \ll \frac{a}{\rho(\tau)} \ll \sqrt{N}
\label{5-8}
\een
Then one gets, using (\ref{2-9})
\ben
S_A  \propto \frac{1}{\pi^2} \left\{ 1+\gamma_E  
 +\log \left[ 4\sqrt{2N}\frac{a}{\rho (\tau)} \right]  \right\}
\label{5-9}
\een
where $\gamma_E$ is Euler's constant. The derivation (\ref{5-9}) is given in Appendix \ref{appB}.

The logarithmic dependence on the subsystem size is characterisic of $1+1$ dimensional systems. For relativistic systems the scale is provided by a UV cutoff. For free non-relativistic fermions on a line the entanglement entropy is finite with the UV cutoff replaced by $N$ \cite{das-ee}. A similar result holds for fermions in an invererted harmonic oscillator potential \cite{das-ee,hartnoll}. For fermions in a harmonic oscillator potential with a constant frequency this logarithmic dependence has been shown in the so called "bulk limit" in \cite{majumdar}.

For CCP protocols, $\rho(\tau)$ oscillates and the condition (\ref{5-8}) continues to hold once it is imposed at early times. However for the ECP $\rho(\tau)$ monotonically increases so that at very late times the condition $\frac{1}{\sqrt{N}} \ll \frac{a}{\rho(\tau)}$ will be violated. It turns out, however, that for the regime
\ben
\frac{a}{\rho(\tau)} \ll \frac{1}{\sqrt{N}}
\label{5-10}
\een
one can use a different approximation which yields
\ben
S_A \propto \frac{\sqrt{N}}{\pi} \frac{a}{\rho(\tau)}
\label{5-11}
\een
Note that the entanglement entropy is now proportional to $a$. However the proportionality constant decreases steadily as $\frac{1}{\tau}$ since the function $\rho(\tau) \sim \tau$ at late times. The derivation of (\ref{5-11}) is given in Appendix \ref{appB}.

\begin{figure}
\centering
\subfloat[$\omega_0 \delta t=0.1$(CCP)]{
\includegraphics[width=0.45\textwidth]{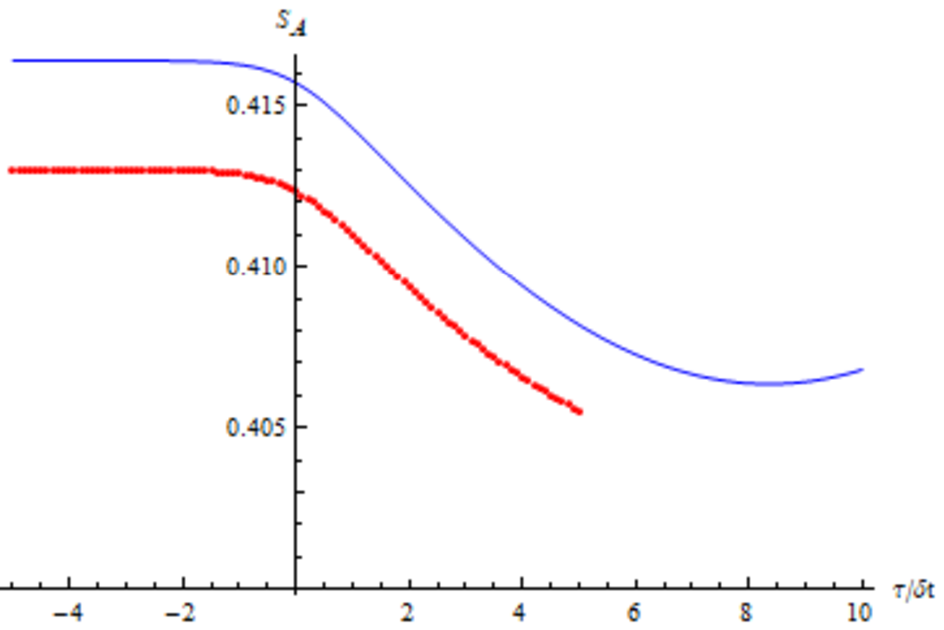}}
\subfloat[$\omega_0 \delta t=0.1$(ECP)]{
\includegraphics[width=0.45\textwidth]{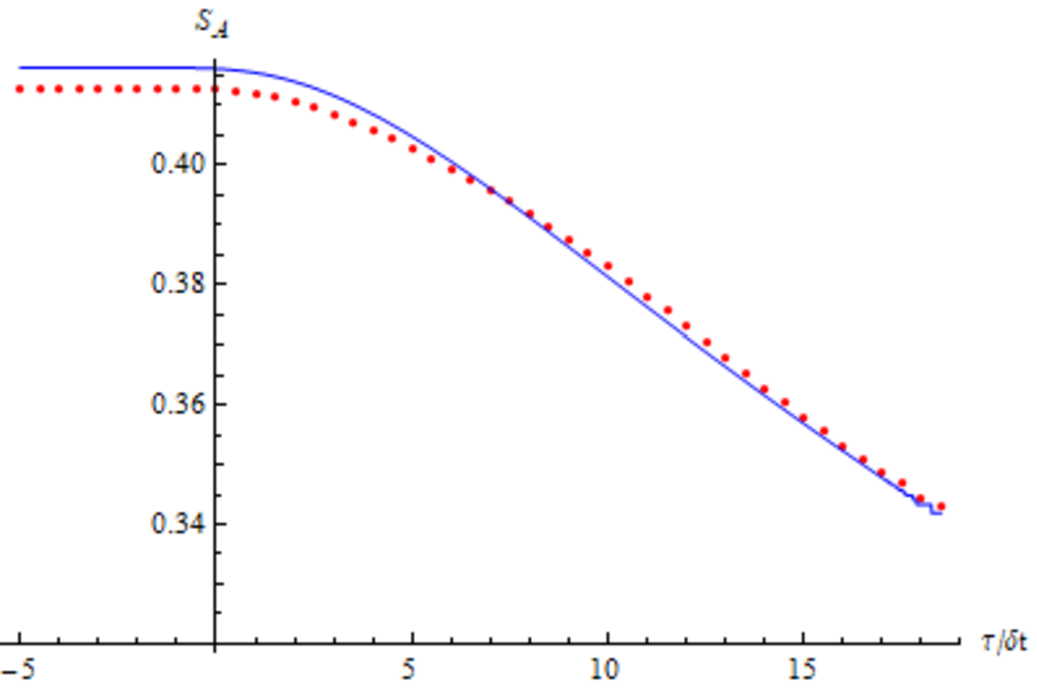}} \\
\subfloat[$\omega_0 \delta t=1$(CCP)]{
\includegraphics[width=0.45\textwidth]{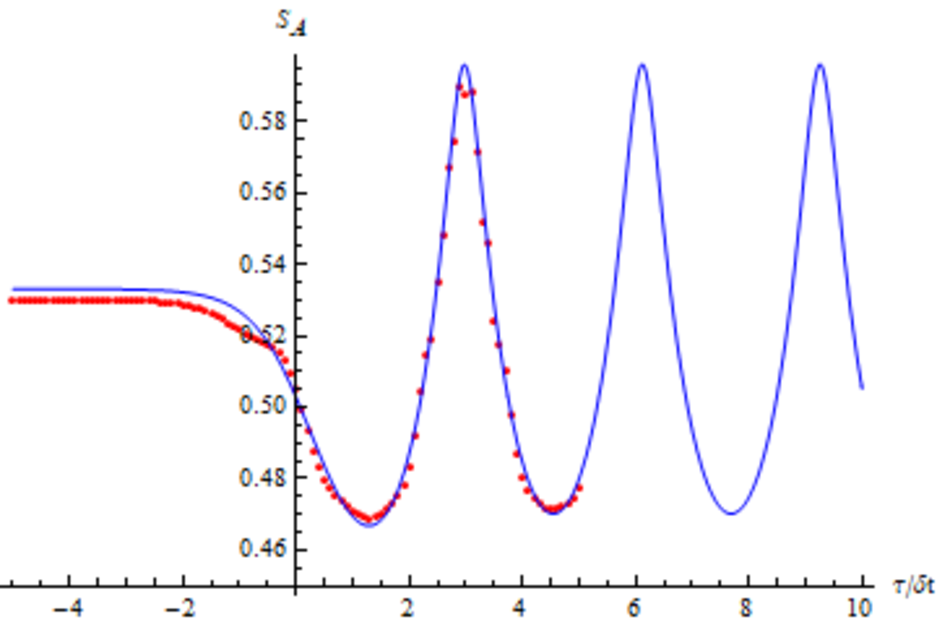}}
\subfloat[$\omega_0 \delta t=1$(ECP)]{
\includegraphics[width=0.45\textwidth]{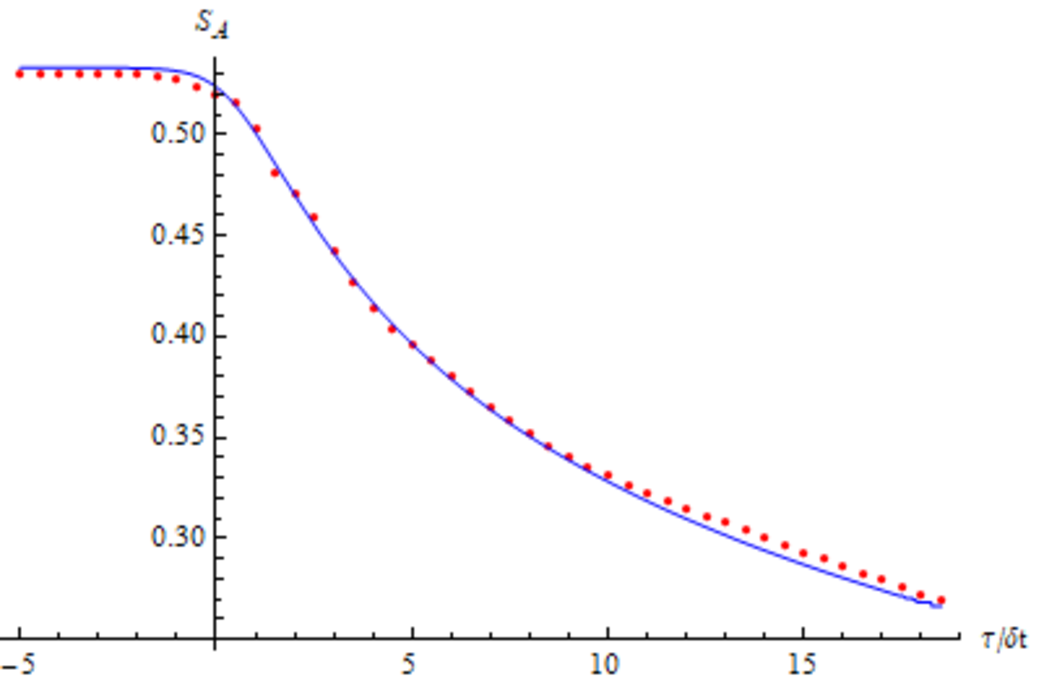}} \\
\subfloat[$\omega_0 \delta t=10$(CCP)]{
\includegraphics[width=0.45\textwidth]{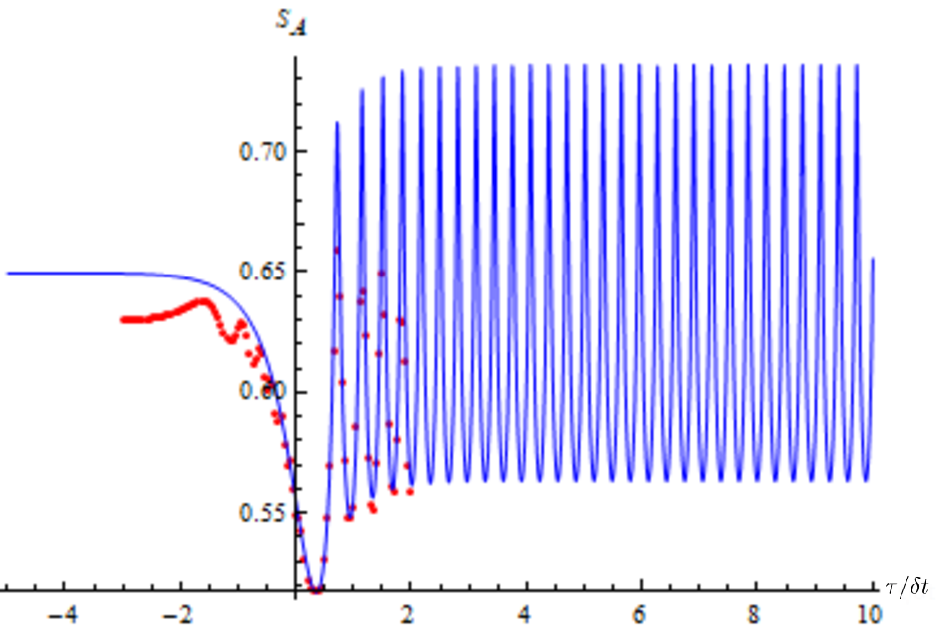}}
\subfloat[$\omega_0 \delta t=10$(ECP)]{
\includegraphics[width=0.45\textwidth]{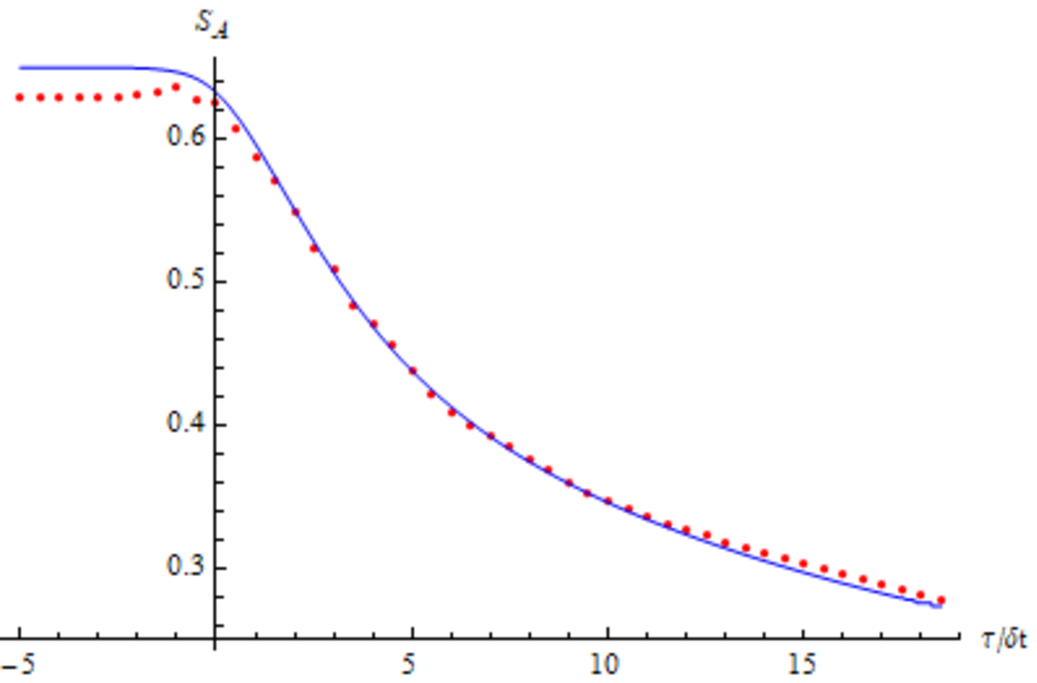}} \\
\caption[Time evolution of EE in various cases.]{Time evolution of Entanglement Entropy $S_A(\tau)$ in various cases. Red dots are exact large N result from (\ref{cxy_ex}). Blue solid lines are results (\ref{5-9}) in the regime (\ref{5-8}). $N=50$, $a=1$.}
\label{fig-three}
\end{figure}

Plots of the time dependence of the entanglement entropy in various cases are shown in Figure \ref{fig-three}. For the cis-critical protocol, the function $\rho(\tau)$ oscillates after an initial increase, so that the effective size of the interval in the equivalent constant frequency problem also oscillates. This would lead to oscillations in the entanglement entropy as well.

For ECP, however, $\rho(\tau)$ decreases continuously. 
This means that for a given $a$ the effective value of the interval in the equivalent constant frequency problem keeps decreasing with time. This should also mean that the entanglement entropy keeps desceasing with time. This is basically because as the fermions are released from the trap they simply spread out : both $\langle N_A \rangle$ and $\langle (\Delta N_A)^2 \rangle$ keep decreasing leading to a loss of entanglement. It follows from (\ref{5-11}) that at late times the entanglement entropy goes to zero as a power law $\sim \frac{1}{\tau}$.

\section{Phase Space Density for Harmonic Oscillator Potential}
\label{sec-nine}
In this section we present the time evolution of the Wigner Distribution function, also called the phase space density, $u(x,p,\tau)$, under CCP and ECP quench protocols in a right side up harmonic oscillator potential. In the classical limit, which is given by $\hbar\to0,N\to \infty$ with $N\hbar = $  fixed, $u(x,p,\tau)$ can only take values of $0$, $1$ since no two fermions can occupy the same position and momentum. A value of $1$ corresponds to the presence of one fermion within a phase space volume between $q$ and $q+dq$ and $p$ and $p + dp$.

Using the canonical transformations in (\ref{1-17}), we can write the time evolved phase space density as a function of the original coordinates as 

\bea
u\bigg({q\over\rho},p\rho - q\dot \rho ,\tau\bigg) = \theta \bigg( 2E_f -  \bigg( ({q\over\rho})^{2} + (p\rho - q\dot \rho)^2 \bigg)\bigg)
\label{u}
\eea
Here $E_f$
is the fermi level and defines the boundary of the phase space density, i.e. the fermi surface. $\theta$ is the Heaviside step function which satisfies the relations
\bea
\theta(x) &=&\bigg\{ \begin{array}{c} 1,\qquad x \geq0\\
 0, \qquad x<0 
 \end{array}
\eea 
Equation (\ref{u}) takes a value of $1$ for $q,p$ which satisfy the relation $ 2E_f \geq   ({q\over\rho})^{2} + (p\rho - q\dot \rho)^2$. This will produce what we call a phase space `droplet'. As time evolves, the shape of this `droplet' will evolve according to the chosen quench protocol. We present the results for the ECP and CCP cases.

\subsection{ECP case}
Here we discuss the time evolution of (\ref{u}) for the ECP case that has a $\rho$ which is given in (\ref{1-14-6}) and (\ref{4-2}).

\begin{figure}
\centering
\subfloat[ $\tau=-2$]{
\includegraphics[width=0.42\textwidth]{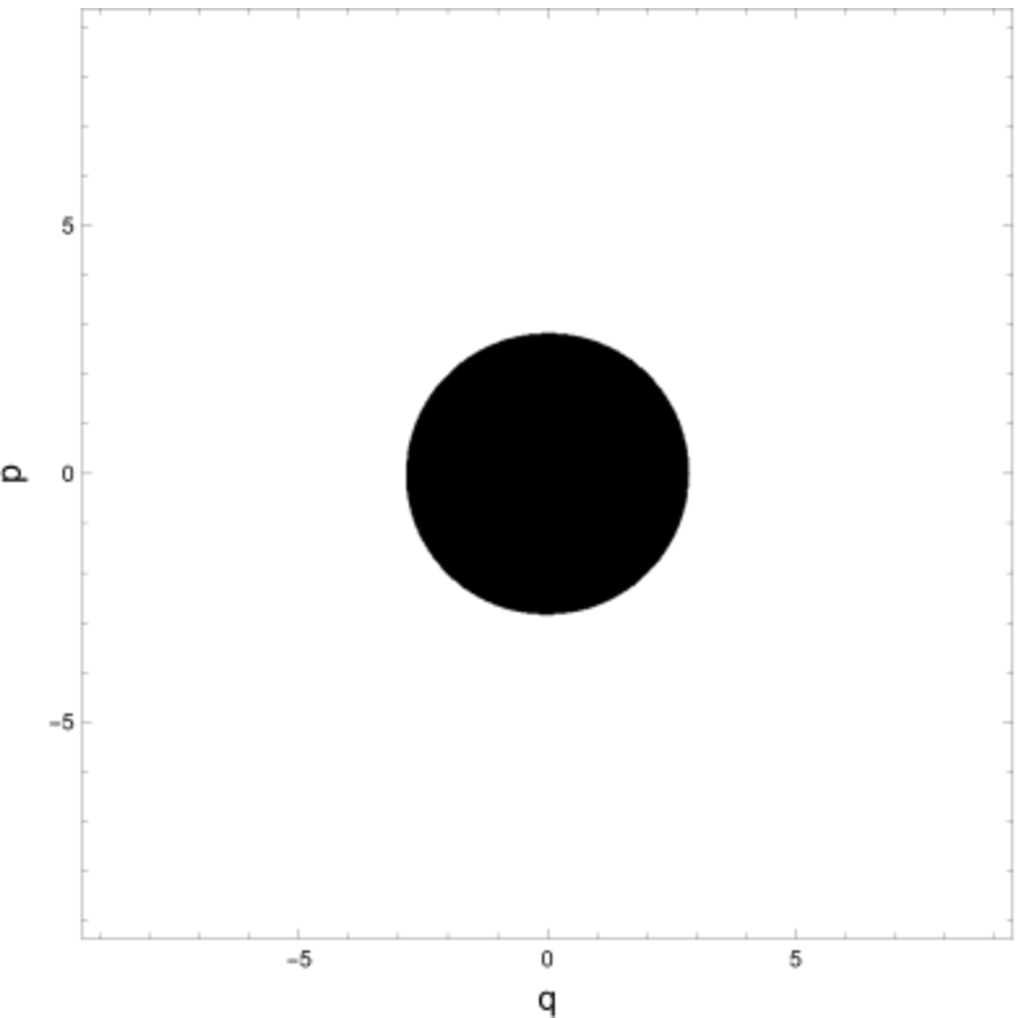}}
\subfloat[$\tau=0$]{
\includegraphics[width=0.42\textwidth]{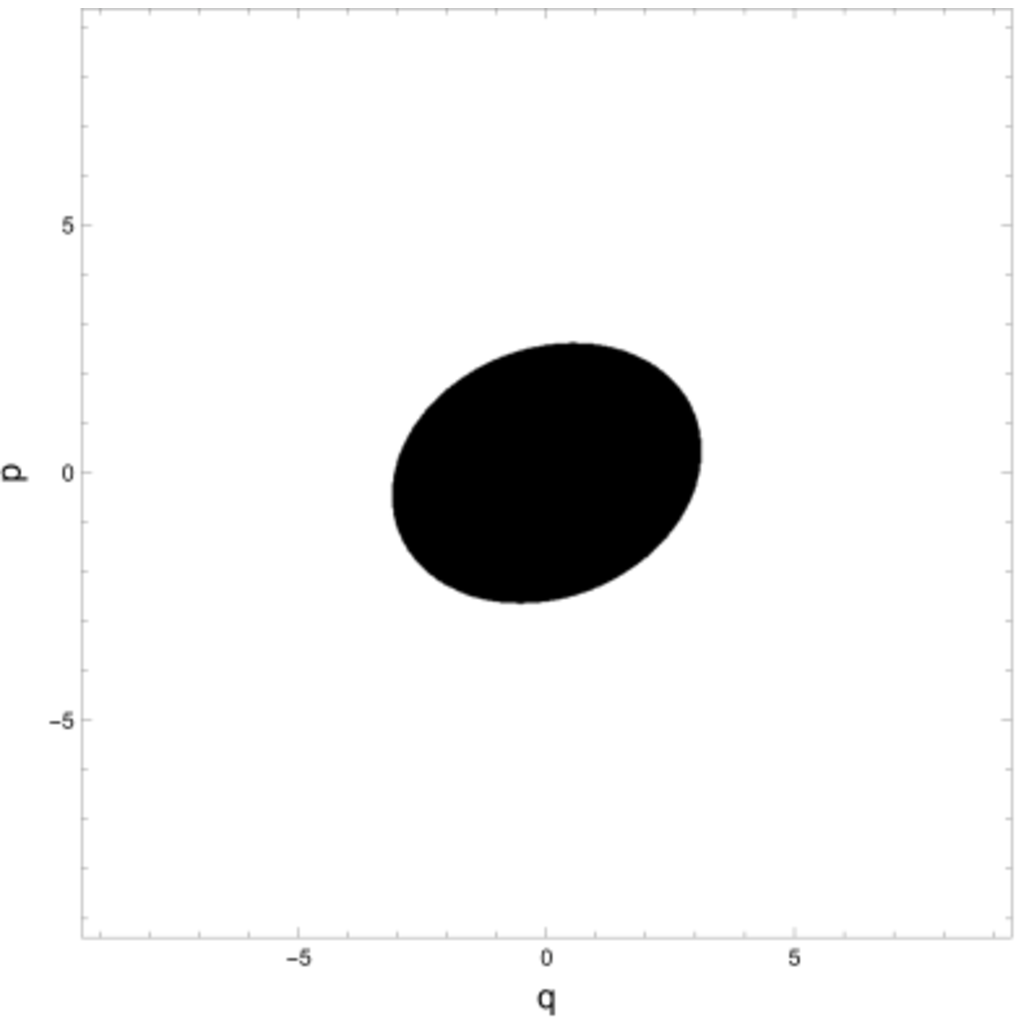}} \\
\subfloat[$\tau=2$]{
\includegraphics[width=0.42\textwidth]{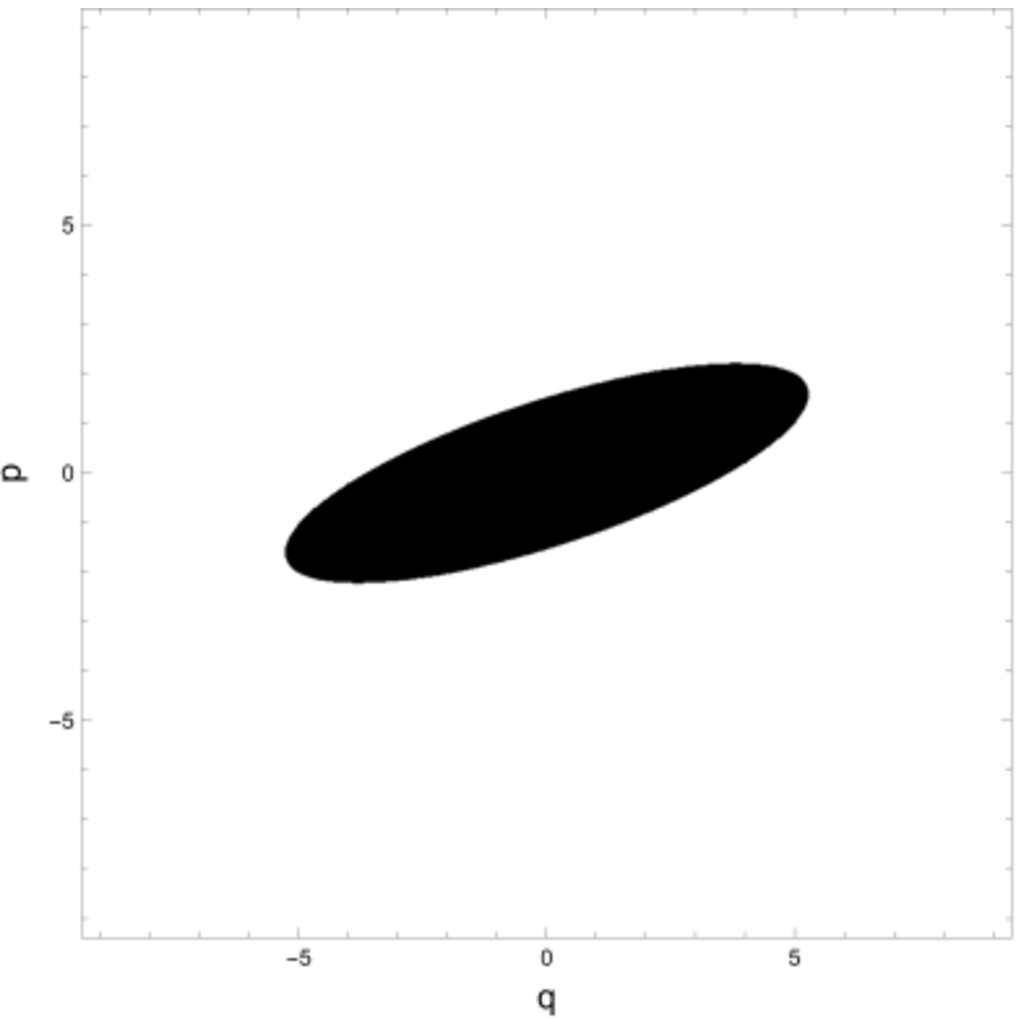}}
\subfloat[$\tau=4$]{
\includegraphics[width=0.42\textwidth]{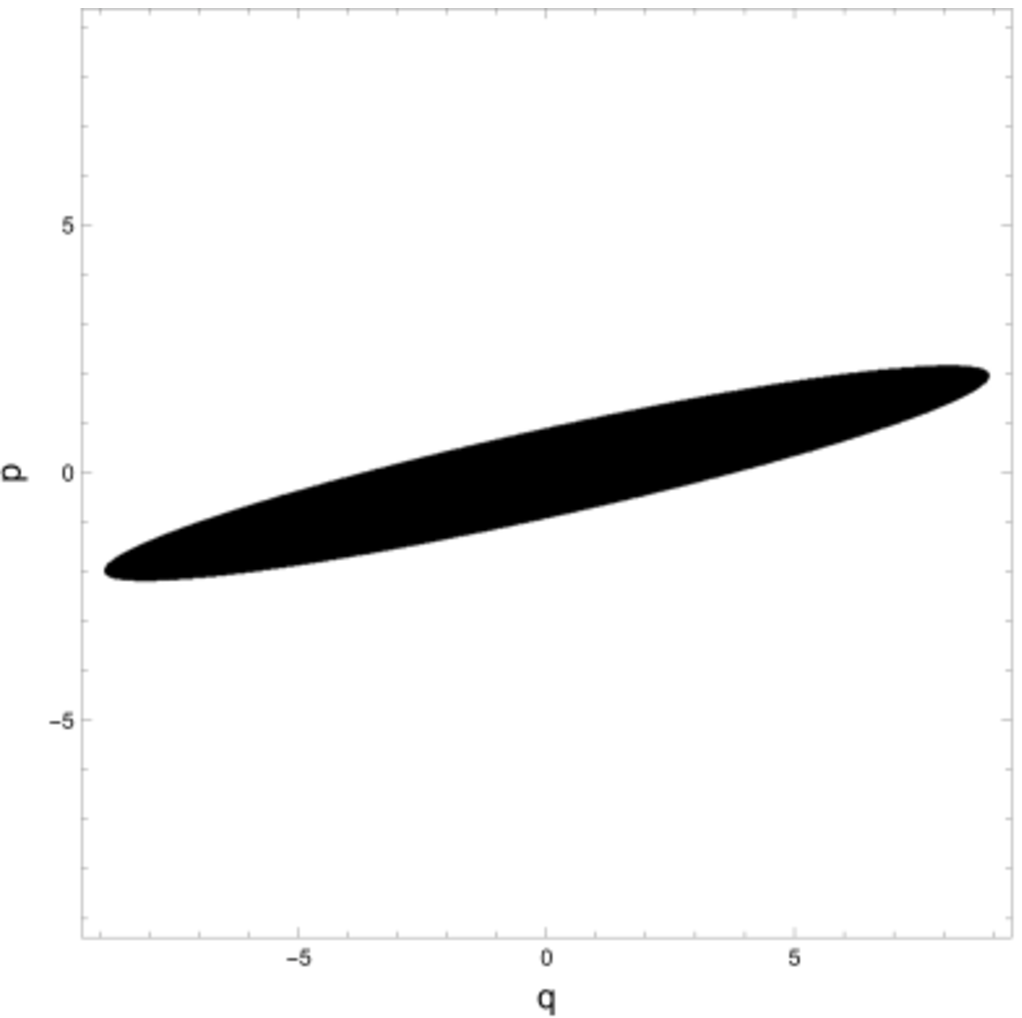}} \\
\caption[Time evolution of Wigner Distribution function.]{Time evolution of a contour plot of the Wigner Distribution function in the classical limit for the ECP case. The black region corresponds to $u=1$ and the white region corresponds to $u=0$. We have taken $\delta t =1 $ , $\omega_0 = 1$. The radius of the initial droplet is $\sqrt{2E_f} = 2\sqrt{2}$ and the area, which is conserved in time, is $N\hbar = 2\pi E_f $.}
\label{fig-uecp}
\end{figure}
In Figure \ref{fig-uecp} we see that the phase space `droplet' spreads out in the upper right and lower left quadrants. This corresponds to motion along both directions of the infinite line. Since we are quenching to zero potential, we are `freeing' the fermions from the harmonic trap and they begin to spread over the real line. The rate at which the `droplet' spreads is related to $\delta t$, the timescale of the quench protocol.

\subsection{CCP case}
Here we discuss the time evolution of (\ref{u}) for the CCP case that has a $\rho$ which is given in (\ref{1-14-6}) and (\ref{3-2}). 

\begin{figure}
\centering
\subfloat[ $\tau=-2$]{
\includegraphics[width=0.42\textwidth]{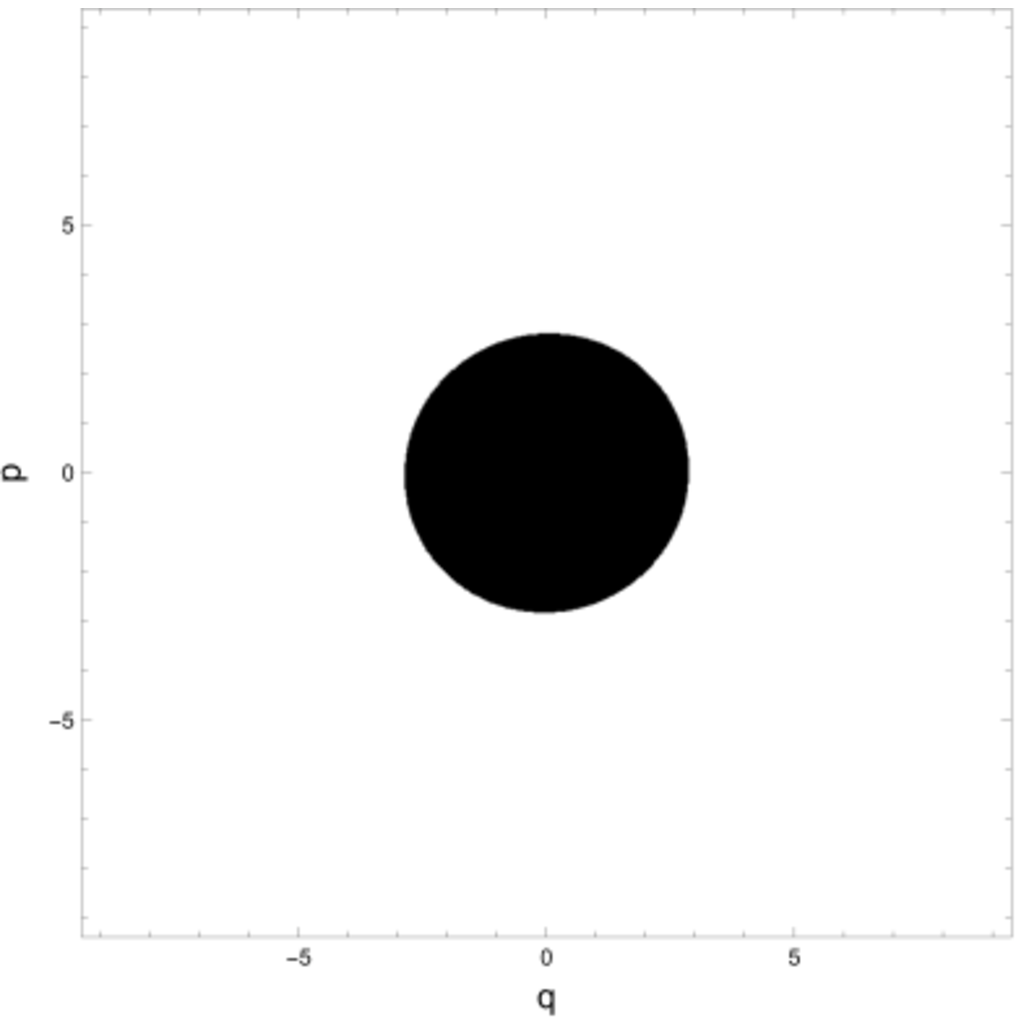}}
\subfloat[$\tau=0$]{
\includegraphics[width=0.42\textwidth]{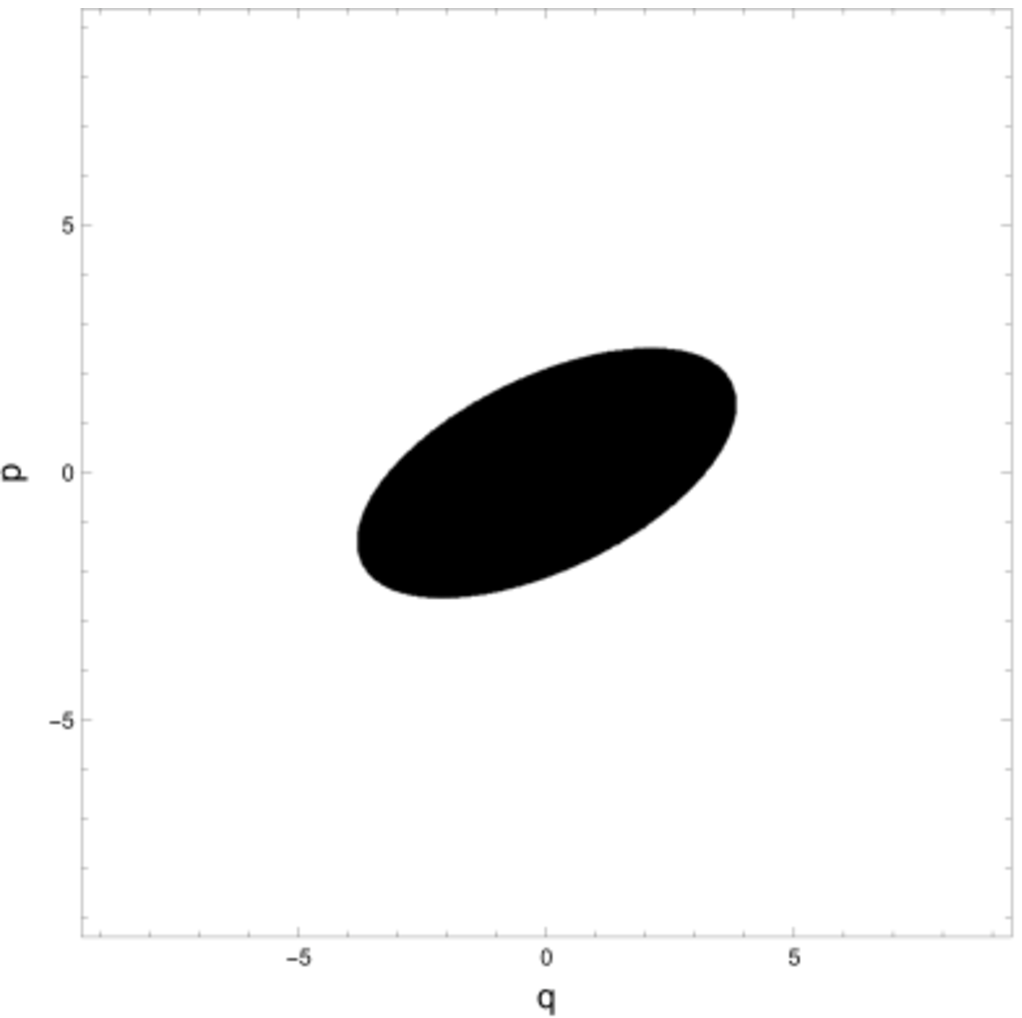}} \\
\subfloat[$\tau=2$]{
\includegraphics[width=0.42\textwidth]{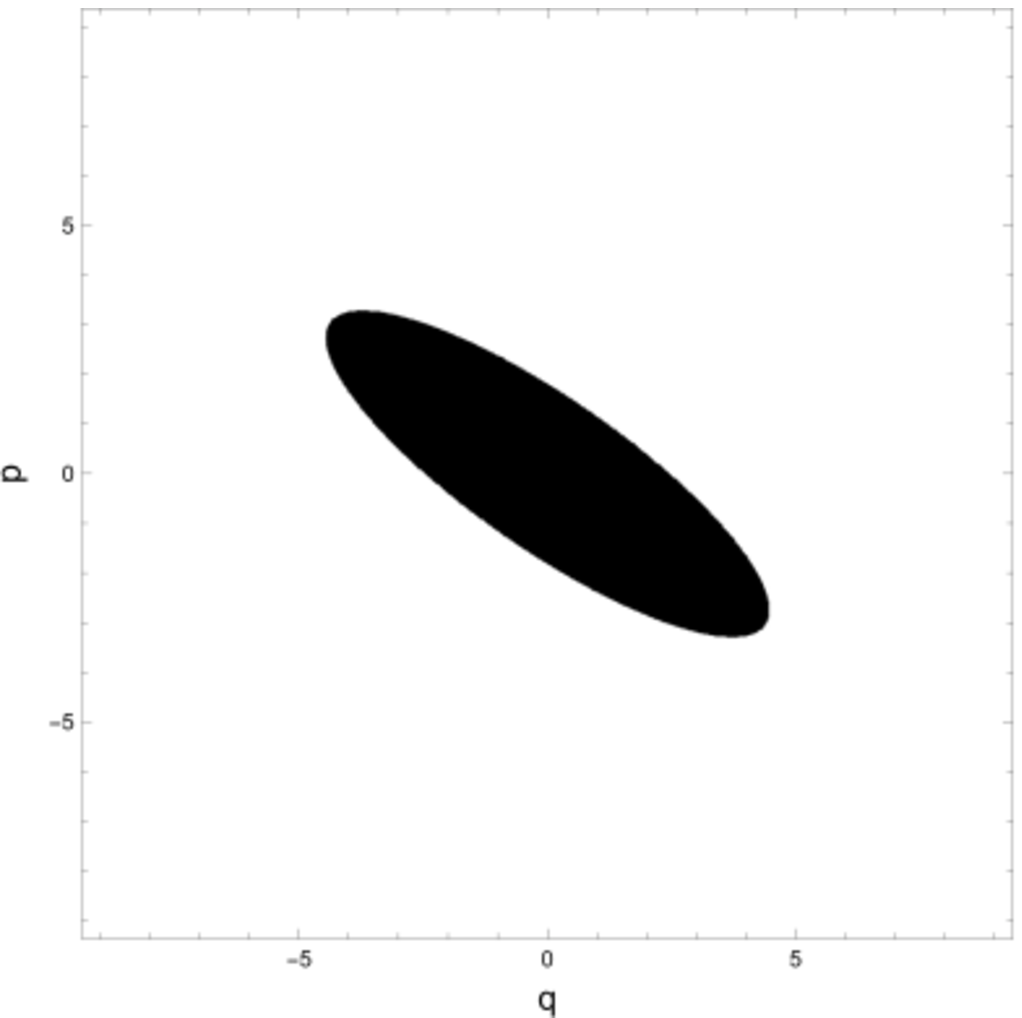}}
\subfloat[$\tau=4$]{
\includegraphics[width=0.42\textwidth]{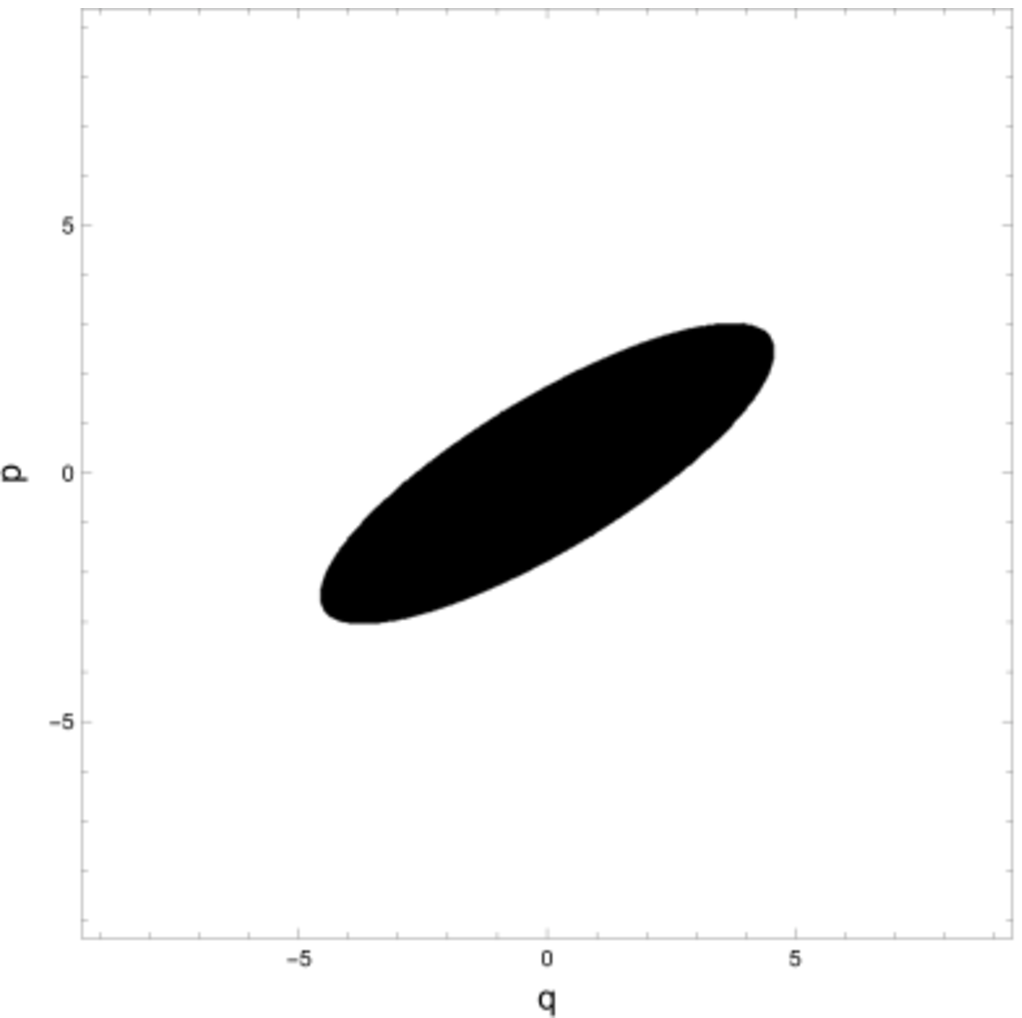}} \\
\caption[Time evolution of Wigner Distribution function.]{Time evolution of the Wigner Distribution function in the classical limit for the CCP case. We have taken $\delta t =1 $ , $\omega_0 = 1$. The radius of the initial droplet is $\sqrt{2E_f} = 2\sqrt{2}$ and the area, which is conserved in time, is $N\hbar = 2\pi E_f $.}
\label{fig-uccp}
\end{figure}

In Figure \ref{fig-uccp} we see that the phase space `droplet' initially spreads out and then begins to rotate in a clockwise fashion. This rotation comes from the oscillatory nature of $\rho$ in the CCP case for $\tau>0$. We can understand the physical origin of this rotation. We are quenching from a potential of frequency $
\omega_0$, to 0, back to $\omega_0$ over a timescale of $\delta t$. The fermions initially just spread along the real line as the potential barrier goes to zero just as in the ECP case. However, when the barrier is restored to its original value, the fermions hit the edge of the restored barrier and then reflect back. This reflection is indicated by the rotation of the stretched droplet in a clockwise fashion. As time evolves the stretched droplet will continue to rotate indefinitely as the electrons keep reflecting off the walls of the potential barrier.

\subsection{Time evolution of perturbations along fermi surface}
\begin{figure}
\centering
\subfloat[ $\tau=0$]{
\includegraphics[width=0.40\textwidth]{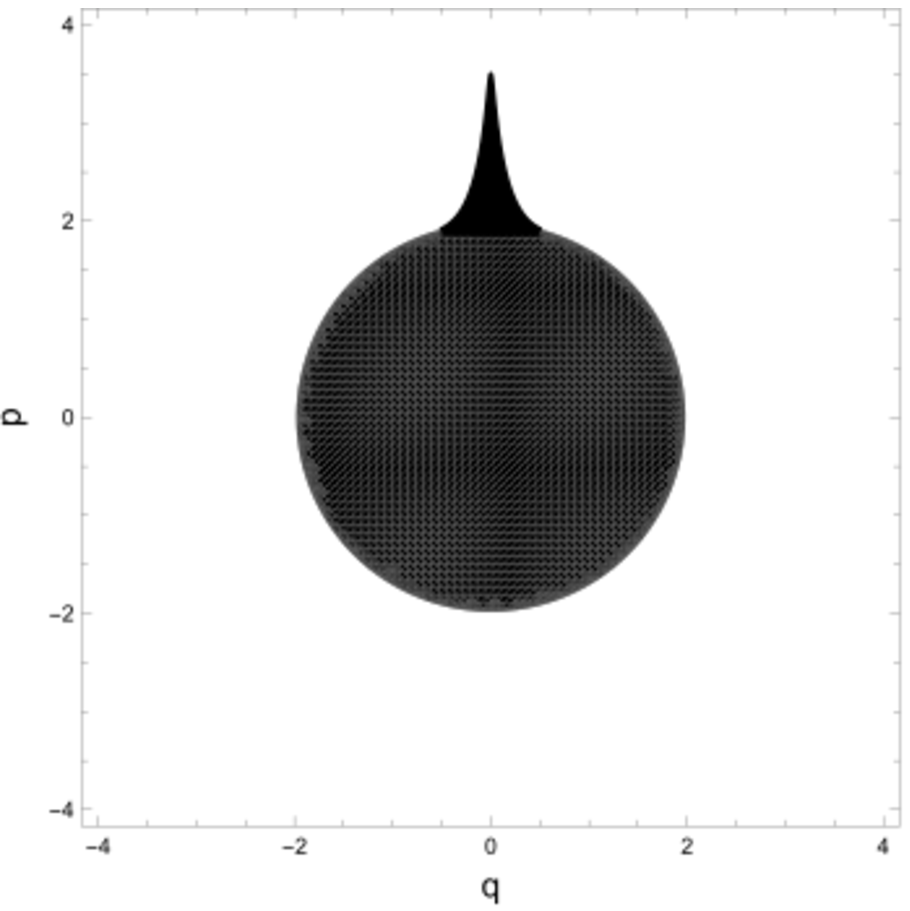}}
\subfloat[$\tau={2\pi\over3}$]{
\includegraphics[width=0.40\textwidth]{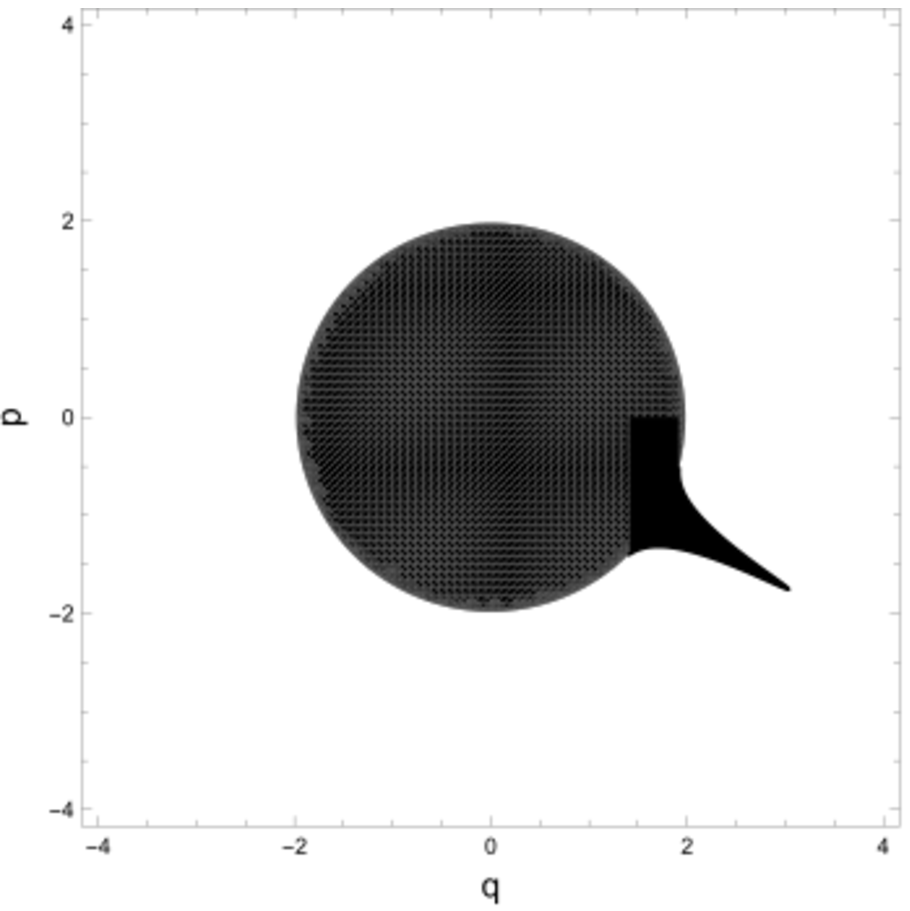}} \\
\subfloat[$\tau={4\pi\over3}$]{
\includegraphics[width=0.40\textwidth]{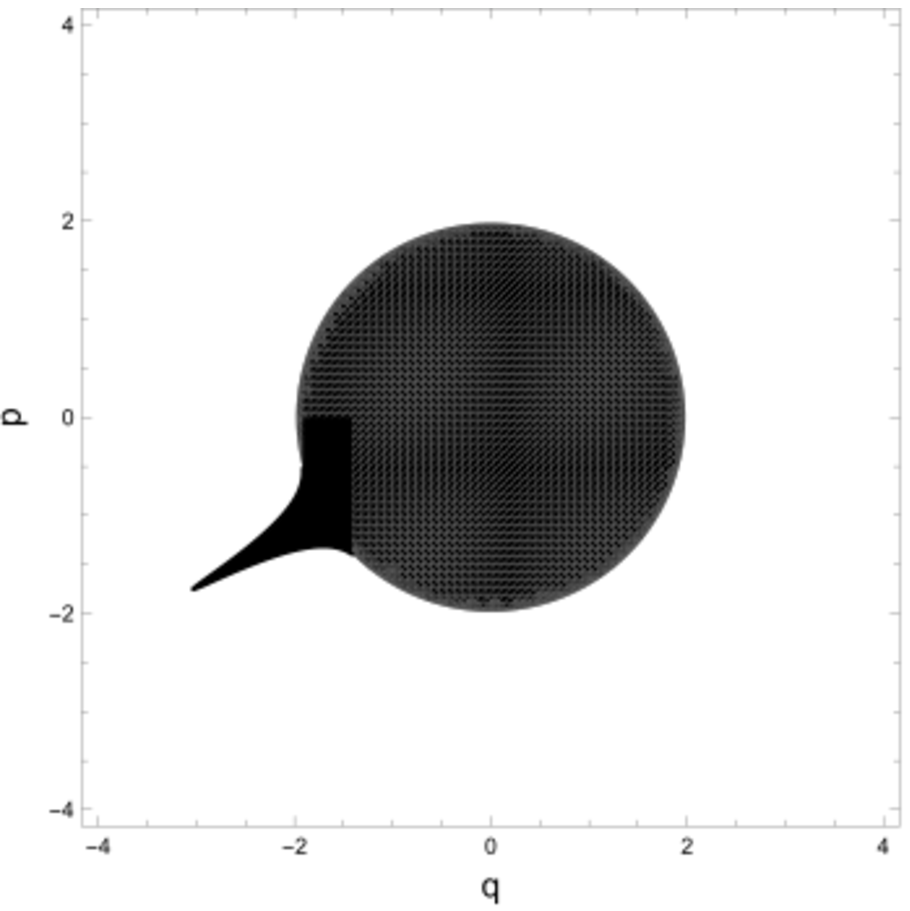}}
\subfloat[$\tau=2\pi$]{
\includegraphics[width=0.40\textwidth]{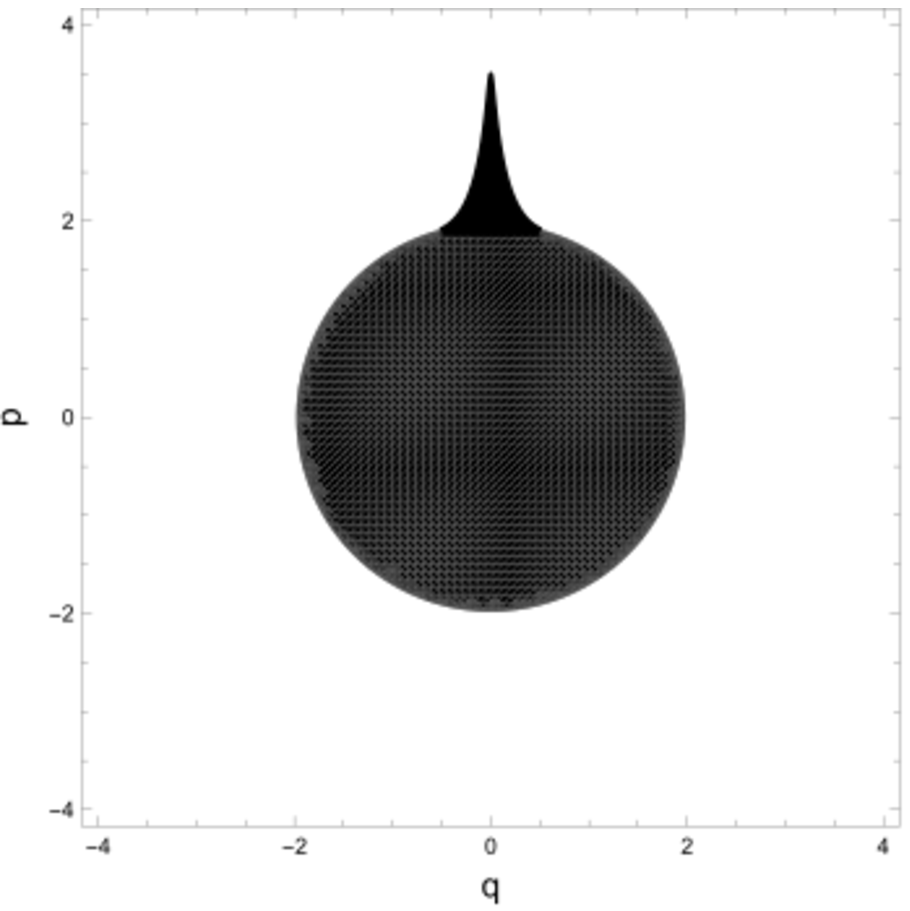}}
\caption[Time evolution of a perturbation of the fermi surface.]{Time evolution of a perturbation of the fermi surface for a time independent harmonic oscillator potential. We have taken $\omega_0 = 1$.}
\label{fig-cf}
\end{figure}

In the previous subsection, we demonstrated the time evolution of a phase space `droplet' under the influence of a right side up harmonic oscillator potential with a time dependent frequency. In this subsection we consider the time evolution of a perturbation of the fermi surface of this `droplet'. We would like to know how this perturbation evolves in time. To gain a better understanding of what happens in this case, let us first consider the time evolution under a harmonic oscillator potential with a time independent frequency. In figure \ref{fig-cf} we plot this evolution. As expected, we find that the perturbation maintains its shape throughout all of time. This is a consequence of the harmonic oscillator frequency being time independent. As a result, all points of an initial perturbation of the fermi surface will move at the same angular frequency for all subsequent times leaving its shape unaltered.

\begin{figure}
\centering
\subfloat[ $\tau=-5$]{
\includegraphics[width=0.40\textwidth]{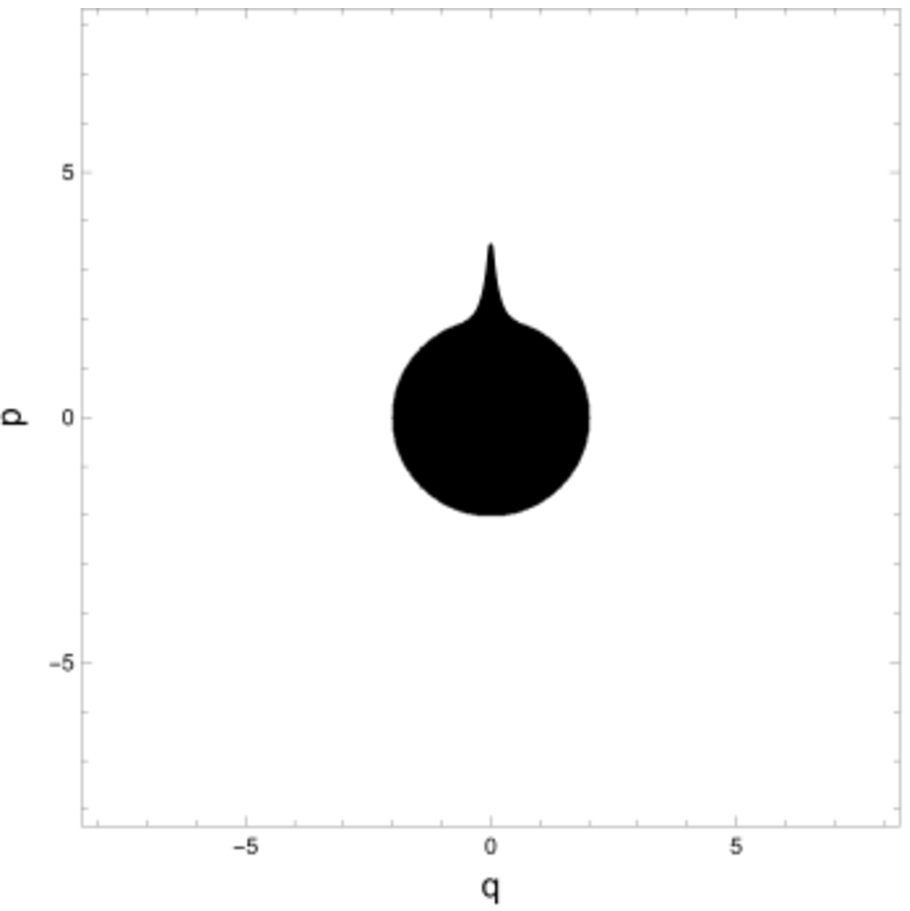}}
\subfloat[$\tau=-3$]{
\includegraphics[width=0.40\textwidth]{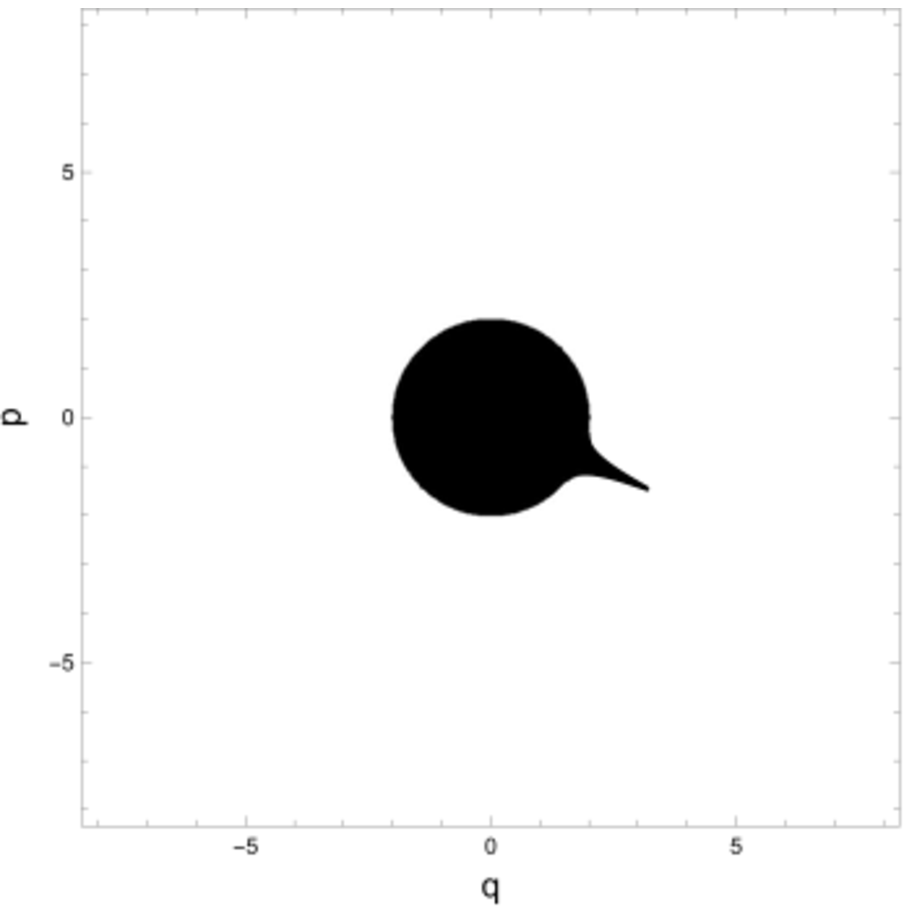}} \\
\subfloat[$\tau=0$]{
\includegraphics[width=0.40\textwidth]{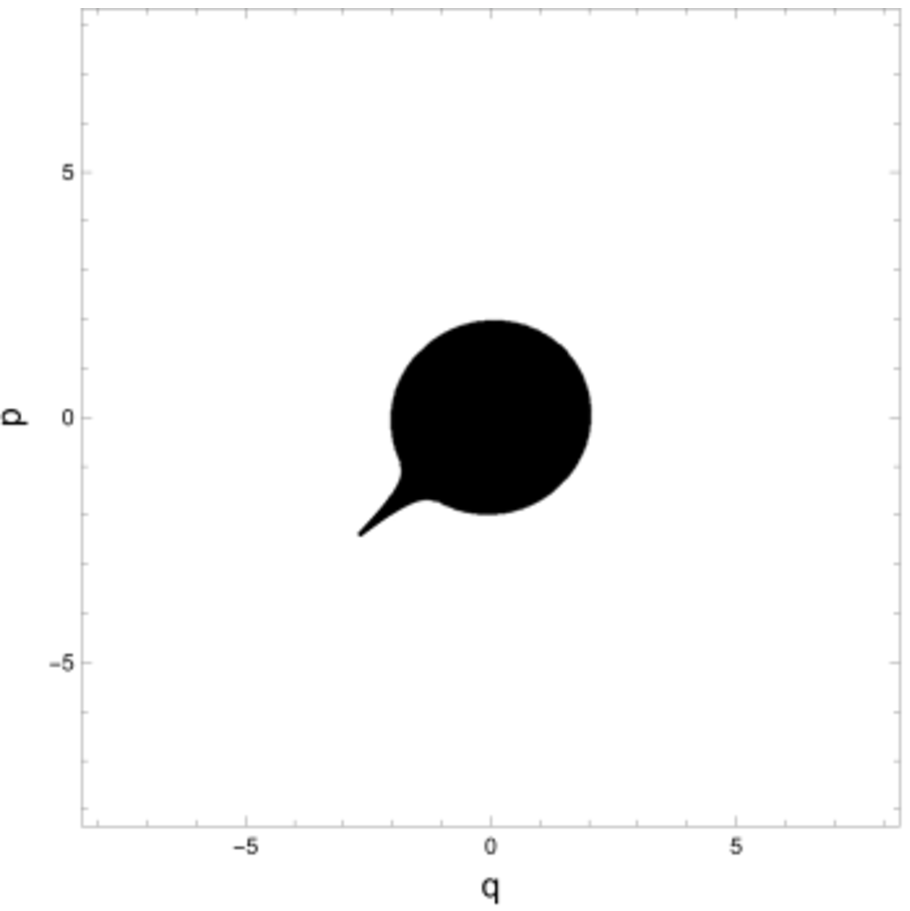}}
\subfloat[$\tau=3$]{
\includegraphics[width=0.40\textwidth]{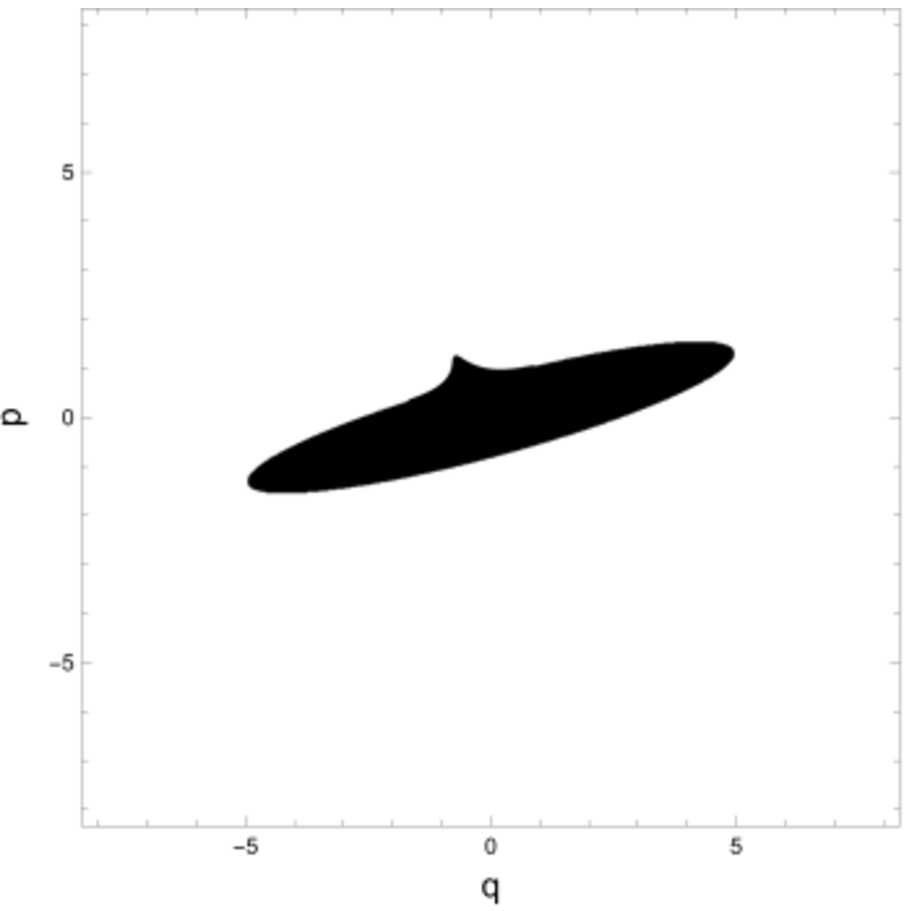}}\\ 
\subfloat[$\tau=4$]{
\includegraphics[width=0.40\textwidth]{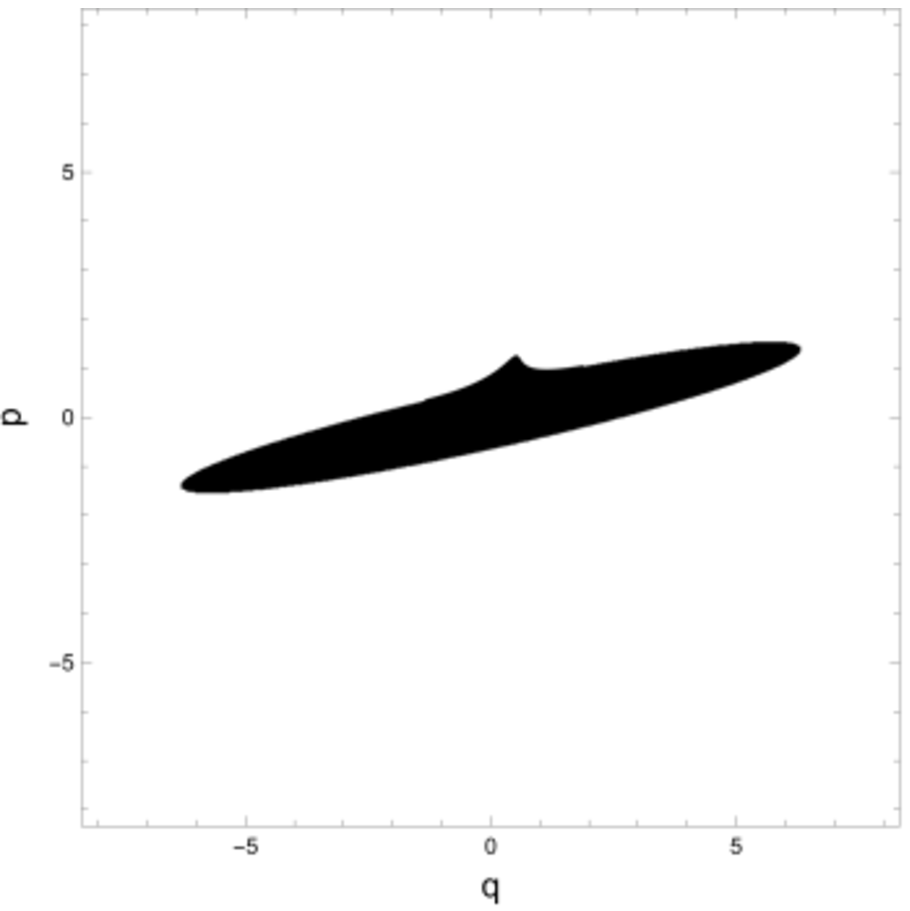}} 
\subfloat[$\tau=5$]{
\includegraphics[width=0.40\textwidth]{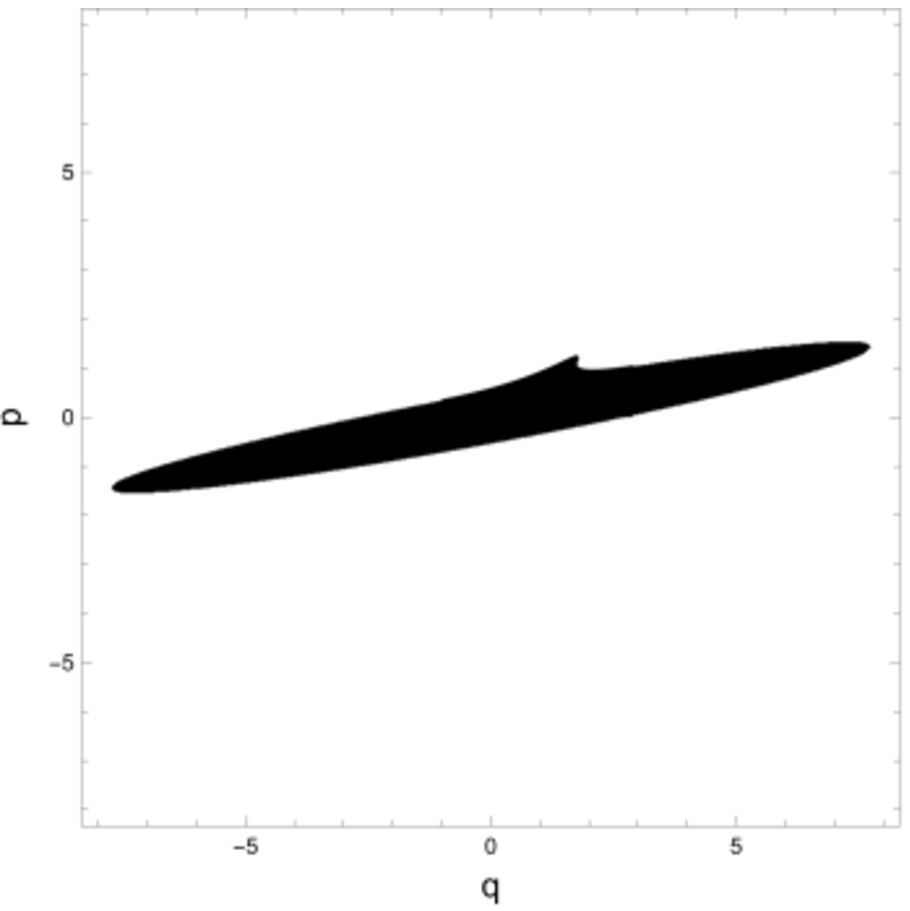}} 
\caption[Time evolution of a perturbation of the fermi surface.]{Time evolution of a perturbation of the fermi surface for the ECP case. We have taken $\delta t =1 $ , $\omega_0 = 1$.}
\label{fig-ecpf}
\end{figure}

\begin{figure}
\centering
\subfloat[ $\tau=3$]{
\includegraphics[width=0.40\textwidth]{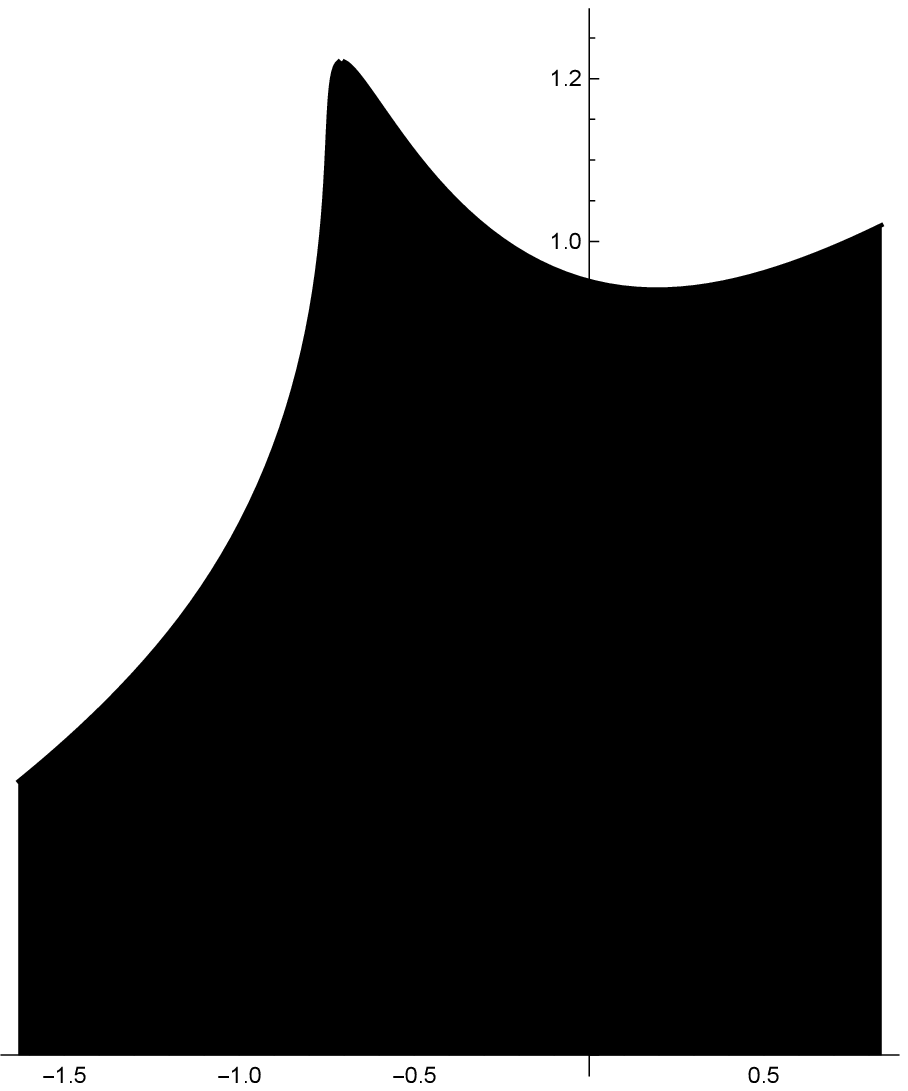}}
\subfloat[$\tau=4$]{
\includegraphics[width=0.40\textwidth]{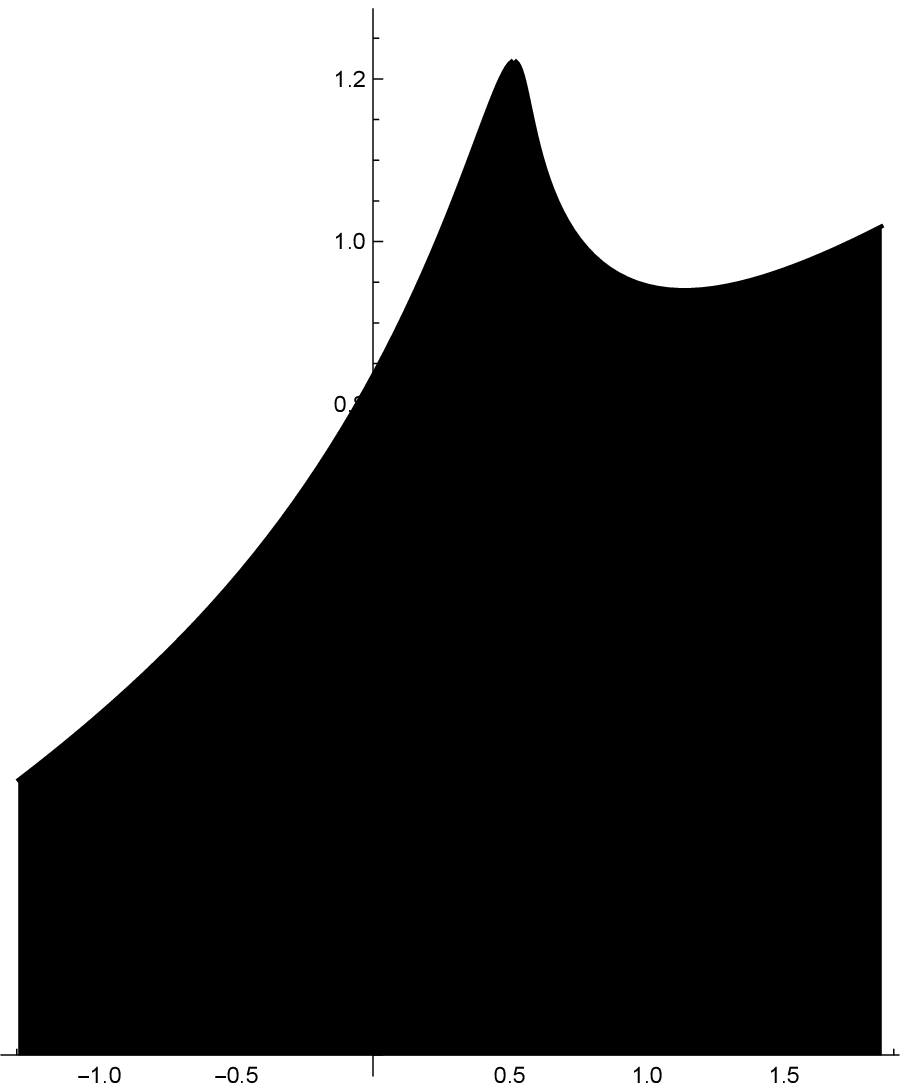}} \\
\subfloat[$\tau=5$]{
\includegraphics[width=0.40\textwidth]{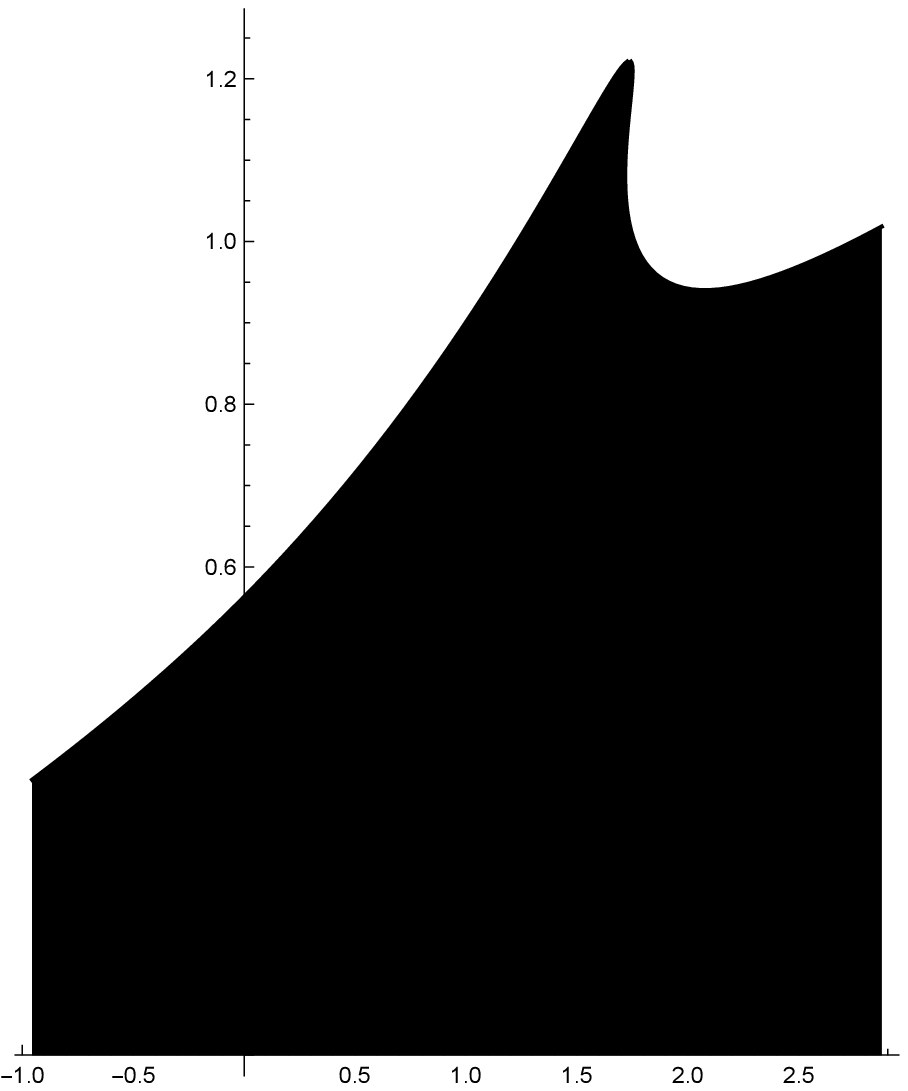}}
\caption[Time evolution of a perturbation of the fermi surface.]{We zoom in to the region of the perturbation of the fermi surface to clearly see a `fold' forming as time evolves.}
\label{fig-ecpfz}
\end{figure}

Now consider the case where an initial perturbation of the fermi surface of a phase space `droplet' evolves under the influence of a right side up harmonic oscillator potential with a time dependent frequency. In particular, we consider the ECP quench protocol. We plot this evolution in Figure's \ref{fig-ecpf}, \ref{fig-ecpfz}. In this case we find something quite interesting. We see that the perturbation develops what we call a `fold'. This is a phenomenon in which a phase space point which is further from the fermi surface moves faster than a phase space point which is closer to the fermi surface. As a result, at some time later than the initial time, the outer most phase space points begin `folding' over towards the fermi surface.

The feature of an initial perturbation developing a fold is characteristic of a system evolving under the influence of a harmonic oscillator potential with any arbitrary time dependent frequency. One can rewrite the phase space coordinates $q,p$ in terms of polar coordinates $r,\theta$. One can then show that ${d\theta\over d\tau}\propto f(\theta,\omega(\tau),\dot \omega (\tau))$ and is therefore not constant in time. On the contrary, if ${d\theta \over d\tau}=\text{const}$, then all the phase space points rotate with the same angular frequency. This is exactly the case for the harmonic oscillator potential with a time independent frequency.

\clearpage

\section{Discussion}

In this paper we considered quantum quench in a nonrelativistic field theory of fermions in an external harmonic oscillator or an invererted harmonic oscillator potential with time dependent mass and frequency. While the strategy we outlined to obtain exact solutions hold for both these potentials, we gave results for the right side harmonic potential in this paper. Explicit solutions for the inverted oscillator potential, which corresponds to quantum quench in the Matrix Model description of two dimensional string theory, will be presented in a future publication \cite{future}.

We examined scaling behavior of observables in the slow and fast quench regime. We found that the slow quench scaling is consistent with Kibble Zurek, and the fast quench scaling is a result of perturbation theory. This system is, however, not suitable to explore if there is a universal fast quench scaling. For the latter we would need to examine a translationally invariant system with an upper bound on the energy spectrum (for example a lattice system) so that a Lieb Robinson bound is possible. We are currently investigating the quench problem in situations like this.


\section*{Acknowledgements}
We would like to thank Gautam Mandal for many discussions and insights, and E. Kiritsis for a conversation. The work of S.R.D and S.L are partially supported by National Science Foundation grants NSF-PHY/1521045 and NSF/PHY-1818878. The work of S.H is supported by the Lyman T. Johnson postdoctoral fellowship.

\appendix
\section{Approximation of $\rho(\tau)^2$ in various limits} \label{appA}

In this appendix we explicitly derive approximated $\rho(\tau)^2$ and therefore $\langle \cO \rangle$ in various limits from the exact CCP solution (\ref{3-2}) and ECP solution (\ref{4-2}). In appendix \ref{A-1} we study the CCP case and in appendix \ref{A-2} we study the ECP case.

\subsection{In CCP}
\label{A-1}

\paragraph{Slow quench ($\omega_0 \delta t \gg 1$)}
We consider the behavior at $\tau=0$, in which case (\ref{3-2}) can be simplified into 
\begin{equation}
f(\tau=0) = \frac{1}{\sqrt{2\omega_0}}  \frac{2^{i\omega_0\delta t}}{E_{1/2}\tilde{E}'_{3/2}-E'_{1/2}\tilde{E}_{3/2}} \tilde{E}'_{3/2}. 
\label{a-1}
\end{equation}
Notice that in (\ref{3-3}), $\text{Re} a = \text{Re} b = \text{Re} \alpha \in [1/4,1/2]$, therefore we can utilize three identities of the Gamma function
\begin{equation}
\Gamma(z)\Gamma(1-z)= \pi \text{csc} \pi z , 0<\text{Re}z <1,
\end{equation} 
and
\begin{equation}
\begin{split}
\Gamma(1+iy)\Gamma(1-iy)= |\Gamma(1+iy)|^2= \frac{\pi y}{\text{sinh}\pi y}, \hfill \\
\Gamma(1/2+iy)\Gamma(1/2-iy)= |\Gamma(1/2+iy)|^2= \frac{\pi }{\text{cosh}\pi y}, \hfill \\
\end{split}
\end{equation}
and simplify the denominator of (\ref{a-1}) into
\begin{equation}
\begin{split}
&\!\!\!\!\!\!\!\!\!\!E_{1/2}\tilde{E}'_{3/2}-E'_{1/2}\tilde{E}_{3/2} \\
&=  \Gamma(1/2)\Gamma(3/2) |\Gamma(i\omega_0 \delta t)|^2 \left( \frac{1}{\Gamma(a+1/2)\Gamma(1/2-a)\Gamma(b)\Gamma(1-b)} - (a \leftrightarrow b) \right) \hfill \\
&=  \frac{1}{2\omega_0 \delta t \text{sinh} \pi \omega_0 \delta t}\left( \text{sin}\pi (1/2-a)\text{sin} (\pi b) - \text{sin}\pi (1/2-b)\text{sin} (\pi a) \right)\\
&= \frac{1}{2\omega_0 \delta t \text{sinh} \pi \omega_0 \delta t}\text{sin} \pi(b-a) \\
&= \frac{i}{2\omega_0 \delta t}.
\end{split}
\label{a-2}
\end{equation}

On the other hand, according to the asymptotic behavior of the Gamma function
\ben
\Gamma(z) \sim \sqrt{2\pi}e^{-z+(z-\frac{1}{2})\log z}, z \to \infty \text{ and } |\arg z|< \pi
\end{equation}
we can find $[\Gamma(z)]^2 \sim \Gamma(z+1/4)\Gamma(z-1/4)$ under the condition. Therefore, the numerator of (\ref{a-1}) satisfies
\begin{equation}
\begin{split}
|\tilde{E}'_{3/2}|^2 
\approx & \frac{\pi}{4} \frac{\pi}{\omega_0 \delta t \text{sinh} \pi \omega_0 \delta t} \left| \frac{1}{\Gamma(1+\frac{i}{4}\sqrt{4\omega_0^2\delta t^2-1} - \frac{i}{2} \omega_0 \delta t)\Gamma(1-\frac{i}{4}\sqrt{4\omega_0^2\delta t^2-1} - \frac{i}{2} \omega_0 \delta t)} \right| \hfill \\ 
& \times  \left| \frac{1}{\Gamma(1/2+\frac{i}{4}\sqrt{4\omega_0^2\delta t^2-1} - \frac{i}{2} \omega_0 \delta t)\Gamma(1/2-\frac{i}{4}\sqrt{4\omega_0^2\delta t^2-1} - \frac{i}{2} \omega_0 \delta t)} \right| \hfill \\
= & \frac{1}{4} \frac{1}{\omega_0 \delta t \text{sinh} \pi \omega_0 \delta t} \left\{ \frac{\text{sinh} \pi (\frac{1}{2}\sqrt{4\omega_0^2\delta t^2-1} - \omega_0 \delta t)}{\frac{1}{2}\sqrt{4\omega_0^2\delta t^2-1} -  \omega_0 \delta t}\frac{\text{sinh} \pi (\frac{1}{2}\sqrt{4\omega_0^2\delta t^2-1} + \omega_0 \delta t)}{\frac{1}{2}\sqrt{4\omega_0^2\delta t^2-1} +  \omega_0 \delta t} \right\}^{1/2} \hfill \\
\to & \frac{1}{2} \frac{1}{\omega_0 \delta t } \left\{ \frac{\pi}{4 \omega_0 \delta t}  \right\}^{1/2} \hfill 
\end{split}
\end{equation}

As a result,
\begin{equation}
\rho^2(\tau=0)=2|f|^2(\tau=0) =2 \frac{1}{{2\omega_0}}  \frac{1}{\frac{1}{(2\omega_0 \delta t)^2}} \frac{1}{2} \frac{1}{\omega_0 \delta t } \left\{ \frac{\pi}{4 \omega_0 \delta t}  \right\}^{1/2} 
= {\sqrt{\pi}}\sqrt{\frac{\delta t}{\omega_0}}
\end{equation}
and thus (\ref{3-4}).

\paragraph{Early time in fast quench ($\omega_0 \tau \ll \omega_0 \delta t \ll 1$)}

When $\omega_0 \tau \ll \omega_0 \delta t \ll 1$, in (\ref{3-3}) $a=b^*$, thus the Hypergeometric functions in (\ref{3-2}) are real. Therefore,
\begin{equation}
 E_c^* = E_c(a \leftrightarrow b) = E'_c,
\label{a8}
\end{equation}
and
\begin{equation}
\begin{split}
\rho^2(\tau) = \frac{1}{{\omega_0}} & \frac{\text{cosh}^{4\alpha}(\tau/\delta t)}{|E_{1/2}\tilde{E}'_{3/2}-E'_{1/2}\tilde{E}_{3/2}|^2}   \times \hfill \\
& \left\{ |\tilde{E}'_{3/2}|^2 {_2F_1}^2(a,b;\frac{1}{2}; -\text{sinh}^2 \frac{\tau}{\delta t}) +| E'_{1/2}|^2 \text{sinh}^2\frac{\tau}{\delta t} {_2F_1}^2(a+\frac{1}{2},b+\frac{1}{2};\frac{3}{2};-\text{sinh}^2\frac{\tau}{\delta t})  \right. \hfill \\
& \left. +\left( \tilde{E}'_{3/2}{E_{1/2}} +{\tilde{E}_{3/2}}E'_{1/2} \right)\text{sinh}\frac{\tau}{\delta t}  {_2F_1}(a,b;\frac{1}{2}; -\text{sinh}^2 \frac{\tau}{\delta t}){_2F_1}(a+\frac{1}{2},b+\frac{1}{2};\frac{3}{2};-\text{sinh}^2\frac{\tau}{\delta t}) \right\}.
\end{split}
\label{a-3}
\end{equation}

Similar to the calculation of (\ref{a-2}), we can find
\begin{equation}
E_{1/2}\tilde{E}_{3/2}' + E_{1/2}'\tilde{E}_{3/2} =\frac{1}{2\omega_0 \delta t \text{sinh} \pi \omega_0 \delta t}\text{sin} \pi(b + a)  \to \frac{1}{2} + \mathcal{O}(\omega_0^4\delta t^4),
\end{equation}
since $\alpha  \sim \frac{1}{2}[1-\omega_0^2\delta t^2]$ when $\omega_0 \delta t \ll 1$.

On the other hand, notice that
\begin{equation}
\Gamma(z+\epsilon) \approx \Gamma(z) + \Gamma'(z) \epsilon + \mathcal{O}(\epsilon^2) = \Gamma(z)(1+\epsilon \psi(z)) +\mathcal{O}(\epsilon^2),
\end{equation}
where $\psi(z) \equiv \Gamma'(z)/\Gamma(z)$ is Digamma function. Moreover, the Gamma function satisfies duplication formula
\begin{equation}
\Gamma(2z)= \frac{1}{\sqrt{2\pi}} 2^{2z-1/2} \Gamma(z) \Gamma(z+1/2).
\end{equation}
Then we can find
\begin{equation}
\begin{split}
\!\!\!\!\!\! |\tilde{E}'_{3/2}|^2&
 \approx  \frac{\pi}{4} |\Gamma(i\omega_0 \delta t)|^2 \left| \frac{1}{\Gamma(1-1/2\omega_0^2\delta t^2 - \frac{i}{2} \omega_0 \delta t)\Gamma(1/2+1/2\omega_0^2\delta t^2- \frac{i}{2} \omega_0 \delta t)} \right|^2 \hfill \\
 \approx & \frac{\pi}{4}  \frac{ |\Gamma(i\omega_0 \delta t)|^2}{|\Gamma(1 - \frac{i}{2} \omega_0 \delta t)|^2|\Gamma(1/2- \frac{i}{2} \omega_0 \delta t)|^2} \frac{1}{\left[ 1-\omega_0^2\delta t^2\text{Re}\psi(1 - \frac{i}{2} \omega_0 \delta t) \right]\left[ 1+\omega_0^2\delta t^2\text{Re}\psi(1/2- \frac{i}{2} \omega_0 \delta t) \right]} \hfill \\
\approx  & \frac{ 1}{4\omega_0^2\delta t^2} \frac{1}{ 1-\omega_0^2\delta t^2\text{Re}\psi(\frac{i}{2} \omega_0 \delta t) +\omega_0^2\delta t^2\text{Re}\psi(1/2+ \frac{i}{2} \omega_0 \delta t)} \hfill \\
\end{split}
\label{a-4}
\end{equation}
We can further simplify it since
\begin{equation}
\psi(2z) = \frac{1}{2} \psi(z) +\frac{1}{2} \psi(z+\frac{1}{2}) + \text{log}\,2
\end{equation}
and
\begin{equation}
\begin{split}
\text{Re}\psi(iy) = 1- \gamma - \frac{1}{1+y^2} + \sum_{n=1}^{\infty}(-1)^{n+1}[\zeta(2n+1)-1]y^{2n}, (|y|<2 ) \hfill \\
\to -\gamma +y^2 +(\zeta(3)-1) y^2
= -\gamma + \zeta(3) y^2, (|y| \ll 1)
\end{split}
\label{a-7}
\end{equation}
and obtain 
\begin{equation}
\begin{split}
 |\tilde{E}'_{3/2}|^2
\approx &  \frac{ 1}{4\omega_0^2\delta t^2} \frac{1}{ 1+2\omega_0^2\delta t^2\left[\text{Re}\psi(i\omega_0 \delta t)-\text{Re}\psi( \frac{i}{2} \omega_0 \delta t)-\text{log}2\right]} \hfill \\
\approx &  \frac{ 1}{4\omega_0^2\delta t^2} \left\{ 1+2\text{log}2 \cdot \omega_0^2\delta t^2+\mathcal{O}(\omega_0^4\delta t^4)\right\}. \hfill \\
\end{split}
\end{equation}

Similarly, we can find
\begin{equation}
| E'_{1/2}|^2 \approx  \frac{1}{4}
\left\{ 1- 2 \text{log}2 \cdot \omega_0^2\delta t^2+\mathcal{O}(\omega_0^4\delta t^4) \right\}.
\end{equation}

Inserting the coefficients back into (\ref{a-3}), we keep the results to order $\omega_0^2\delta t^2$ and $\tau/\delta t$, s.t. $ {_2F_1}^2(\tilde{a},\tilde{b};\tilde{c}; -\text{sinh}^2 \frac{\tau}{\delta t}) \sim 1 $ for arbitrary $(a,b,c)$. We find
\begin{equation}
\begin{split}
\rho^2(\tau) =\frac{1}{{\omega_0}} & \text{cosh}^{4\alpha}(\tau/\delta t)   \times \hfill \\
& \left\{ \left\{ 1+2\text{log}2 \cdot \omega_0^2\delta t^2\right\} {_2F_1}^2(a,b;\frac{1}{2}; -\text{sinh}^2 \frac{\tau}{\delta t}) +\omega_0^2 \delta t^2
\text{sinh}^2\frac{\tau}{\delta t} {_2F_1}^2(a+\frac{1}{2},b+\frac{1}{2};\frac{3}{2};-\text{sinh}^2\frac{\tau}{\delta t}) \right.  \hfill \\
& \left. +2\omega_0^2 \delta t^2\text{sinh}\frac{\tau}{\delta t}  {_2F_1}(a,b;\frac{1}{2}; -\text{sinh}^2 \frac{\tau}{\delta t}){_2F_1}(a+\frac{1}{2},b+\frac{1}{2};\frac{3}{2};-\text{sinh}^2\frac{\tau}{\delta t}) +\mathcal{O}(\omega_0^4\delta t^4)\right\} \hfill \\
=\frac{1}{{\omega_0}} & 
\left\{  1+2\text{log}2 \cdot \omega_0^2\delta t^2 
+2\omega_0^2 \delta t \cdot \tau  +\mathcal{O}(\omega_0^4\delta t^4, \frac{\tau^2}{\delta t^2})\right\} \hfill \\
\end{split}
\label{a18}
\end{equation}
and therefore (\ref{3-10}).

\paragraph{Late time in fast quench ($\omega_0 \delta t \ll \omega_0 \tau \ll 1$)}

Rewrite (\ref{3-2}) by applying identity
\begin{equation}
\begin{split}
_2F_1(a,b;c;z) =& \frac{\Gamma(c)\Gamma(b-a)}{\Gamma(b)\Gamma(c-a)}(-z)^{-a} {_2F_1(a,1-c+a;1-b+a;\frac{1}{z})}  \hfill \\
&+ \frac{\Gamma(c)\Gamma(a-b)}{\Gamma(a)\Gamma(c-b)}(-z)^{-b} {_2F_1(b,1-c+b;1-a+b;\frac{1}{z})}.
\end{split}
\label{a19}
\end{equation}
to each Hypergeometric function on the RHS.
When $\tau >0$, $(-z)^{1/2}=\sinh \frac{\tau}{\delta t}$, thus we can find 
\begin{equation}
\begin{split}
f_{CCP}
=& \frac{1}{\sqrt{2\omega_0}}  \frac{2^{i\omega_0\delta t} \text{cosh}^{2\alpha}(\tau/\delta t)}{E_{1/2}\tilde{E}'_{3/2}-E'_{1/2}\tilde{E}_{3/2}}   \times \hfill \\
& \left\{ \left( E_{1/2}\tilde{E}_{3/2}' + E_{1/2}'\tilde{E}_{3/2}\right) (-z)^{-a} {_2F_1(a,1/2+a;1-b+a;\frac{1}{z})} \right. \hfill \\
&\left. + 2E_{1/2}'\tilde{E}_{3/2}' (-z)^{-b} {_2F_1(b+\frac{1}{2},b;1-a+b;\frac{1}{z})} \right\}, \hfill \\
\end{split}
\label{a-5}
\end{equation}
where  $z \equiv -\sinh^2 \frac{\tau}{\delta t}$.

Similar calculations to (\ref{a-4}) show
\begin{equation}
\begin{split}
2E_{1/2}'\tilde{E}_{3/2}' 
=& 2\frac{\pi}{2} \Gamma(-i \omega_0 \delta t)^2 \cdot \frac{1}{\Gamma(a)\Gamma(a+1/2)\Gamma(1/2-b)\Gamma(1-b)} 
\hfill \\
=& 2\frac{\pi}{2} \Gamma(-i \omega_0 \delta t)^2 \left(  \frac{1}{\sqrt{2\pi}} 2^{2a-1/2} \frac{1}{\Gamma(2a)} \cdot  \frac{1}{\sqrt{2\pi}} 2^{2(1/2-b)-1/2}\frac{1}{\Gamma(1-2b)}\right)
\hfill \\
\approx & \frac{1}{2}  \cdot 2^{-i2\omega_0 \delta t} \frac{\Gamma(-i \omega_0 \delta t)^2}{\Gamma(1-i\omega_0 \delta t)\left( 1- \omega_0^2 \delta t^2 \psi(1- i\omega_0 \delta t) \right)\Gamma(-i\omega_0\delta t)\left( 1+ \omega_0^2\delta t^2 \psi(-i \omega_0 \delta t) \right)} \hfill \\
\approx & \frac{i}{2\omega_0 \delta t}  \cdot 2^{-i2\omega_0 \delta t} \frac{1}{\left[ 1+ \omega_0^2\delta t^2\left( \psi(-i \omega_0 \delta t) - \psi(1- i\omega_0 \delta t)\right) \right]} \hfill \\
\end{split}
\end{equation}
We further simplify the equation by using relations
\begin{equation}
\text{Re} \psi(iy) = \text{Re} \psi(-iy) = \text{Re} \psi(1+iy) =\text{Re} \psi(1-iy),
\end{equation}
and
\begin{eqnarray}
\text{Im}\psi(iy)=\frac{1}{2y}+ \frac{1}{2} \pi \text{coth}\pi y, \\
\text{Im}\psi(1+iy)=-\frac{1}{2y}+ \frac{1}{2} \pi \text{coth}\pi y. 
\end{eqnarray}
Then we see that
\begin{equation}
\psi(-i \omega_0 \delta t) - \psi(1- i\omega_0 \delta t) = i\text{Im} \psi(-i \omega_0 \delta t) -i \text{Im}\psi(1- i\omega_0 \delta t) = - \frac{i}{\omega_0 \delta t}
\end{equation}
thus
\begin{equation}
2E_{1/2}'\tilde{E}_{3/2}' =  \frac{i}{2\omega_0 \delta t}  \cdot 2^{-i2\omega_0 \delta t} \frac{1}{1-i \omega_0\delta t }
\end{equation}

Notice that when $\tau \gg \delta t$, $ -\text{sinh}^2 \frac{\tau}{\delta t} \to -\frac{e^{2\tau/\delta t}}{4} + \frac{1}{2}$, $\text{cosh}^2 \frac{\tau}{\delta t} \to \frac{e^{2\tau/\delta t}}{4} + \frac{1}{2} $, and therefore $_2F_1 \to 1 + \mathcal{O}(e^{-2\tau/\delta t})$. Thus, after inserting the coefficients into (\ref{a-5}) and expanding the result to the order $\omega_0 \delta t$, we obtain
\begin{equation}
\begin{split}
f
=  \frac{1}{\sqrt{2\omega_0}}  \text{cosh}^{2\alpha}(\tau/\delta t) &
 \left\{ -i {\omega_0\delta t}2^{i\omega_0\delta t}(-z)^{-a} {_2F_1(a,1/2+a;1-b+a;\frac{1}{z})} \right. \hfill \\
&\left. +  (1+i \omega_0\delta t ) 2^{-i\omega_0 \delta t}  (-z)^{-b} {_2F_1(b+\frac{1}{2},b;1-a+b;\frac{1}{z})} \right\} \hfill \\
\to   \frac{1}{\sqrt{2\omega_0}} \left(\frac{e^{ \tau/\delta t}}{2}\right)^{2\alpha} &
   \left\{ -i {\omega_0\delta t}2^{i\omega_0\delta t}\left(\frac{e^{ \tau/\delta t}}{2}\right)^{-2a} \left( 1 + \mathcal{O}(e^{-2\tau/\delta t}) \right) \right. \hfill \\
 & \left. +  (1+i \omega_0\delta t ) 2^{-i\omega_0 \delta t} \left(\frac{e^{ \tau/\delta t}}{2}\right)^{-2b}\left( 1 + \mathcal{O}(e^{-2 \tau/\delta t}) \right) + \mathcal{O}(\omega_0^2 \delta t^2) \right\} \hfill \\
=    \frac{1}{\sqrt{2\omega_0}} & 
   \left\{ e^{- i\omega_0 \tau} +2\omega_0\delta t \text{sin} \omega_0 \tau  +\mathcal{O}(\omega_0^2 \delta t^2) \right\}. \hfill \\
\end{split}
\end{equation}
i.e.
\begin{equation}
\rho^2(\tau) =2|f|^2 \approx \frac{1}{\omega_0} \left( 1 +2\omega_0\delta t \text{sin} 2\omega_0 \tau  +\mathcal{O}(\omega_0^2 \delta t^2), \right) 
\label{a28}
\end{equation}
and therefore (\ref{3-12}).

\subsection{In ECP}
\label{A-2}
\paragraph{Late-time approximation ($\tau \gg \delta t$, and $\tau \gg \delta t \log \omega_0 \delta t  $)}
According to identity
\begin{equation}
\begin{split}
_2F_1(a,b;a+b;z)= \frac{\Gamma(a+b)}{\Gamma(a)\Gamma(b)}\sum_{n=0}^{\infty}\frac{(a)_n(b)_n}{(n!)^2}[2\psi(n+1) -\psi(a+n)-\psi(b+n)-\log(1-z)](1-z)^n, \hfill \\
(|1-z|<1 \& |\arg(1-z)|<\pi)
\label{a29}
\end{split}
\end{equation}
we can see that for large $\tau/\delta t$, in which case
\begin{equation}
z \equiv \frac{1+\text{tanh}(\tau/\delta t )}{2} = 1- e^{-2\tau/\delta t},
\end{equation}
(\ref{4-2}) can be rewritten into
\begin{equation}
\begin{split}
f_{ECP} \to  \frac{1}{\sqrt{2\omega_{0}}} & \text{exp}\left[-\frac{i}{2}\omega_0 \tau + \frac{i}{2}\omega_0 \delta t \text{log}(e^{\tau/\delta t})\right]
 {_2}F_1 \left[1-\frac{i}{2}\omega_0\delta t, -\frac{i}{2}\omega_0\delta t ; 1-i\omega_{0}\delta t;  1- e^{-2\tau/\delta t} \right]  \hfill \\
= \frac{1}{\sqrt{2\omega_{0}}} & \frac{\Gamma(1-i\omega_{0}\delta t)}{\Gamma(1-\frac{i}{2}\omega_0\delta t)\Gamma(-\frac{i}{2}\omega_0\delta t)} \times  \hfill \\
 \sum_{n=0}^{\infty} & \frac{(1-\frac{i}{2}\omega_0\delta t)_n(-\frac{i}{2}\omega_0\delta t)_n}{(n!)^2} \left[2\psi(n+1) -\psi(1-\frac{i}{2}\omega_0\delta t+n)-\psi(-\frac{i}{2}\omega_0\delta t+n)+2\frac{\tau}{\delta t}\right](e^{-2\tau/\delta t})^n \hfill \\
\end{split}
\label{a-6}
\end{equation}

When  $\tau \gg \delta t \log \omega_0 \delta t  $, one can keep the leading term\footnote{ $\tau \ge \delta t \log \omega_0 \delta t  $ is a sufficient condition to keep (\ref{a-6}) to the leading term.}, then by using Digamma function
\begin{equation}
\psi(1)= -\gamma_E
\end{equation}
and 
\begin{equation}
\psi(1-z)=\psi(z) +\pi \cot \pi z,
\end{equation}
we find
\begin{equation}
f_{ECP} 
\to \frac{1}{\sqrt{2\omega_{0}}}  \frac{\Gamma(1-i\omega_{0}\delta t)}{\Gamma(1-\frac{i}{2}\omega_0\delta t)\Gamma(-\frac{i}{2}\omega_0\delta t)}
\left[-2\gamma_E -2 \Re  \psi \left(\frac{i}{2}\omega_0\delta t \right) + i \pi \coth \frac{\pi\omega_0\delta t}{2}+2\frac{\tau}{\delta t}\right] + \mathcal{O}(e^{-2\tau/\delta t}) 
\end{equation}
Therefore, 
\begin{equation}
\begin{split}
\rho^2(\tau) \sim &\frac{1}{{\omega_{0}}} \left| \frac{\Gamma(1-i\omega_{0}\delta t)}{\Gamma(1-\frac{i}{2}\omega_0\delta t)\Gamma(-\frac{i}{2}\omega_0\delta t)} \right|^2
\left[ \left(-2\gamma_E -2 \Re  \psi \left(\frac{i}{2}\omega_0\delta t \right)+2\frac{\tau}{\delta t} \right)^2 + \left( \pi \coth \frac{\pi\omega_0\delta t}{2}\right)^2 \right] \hfill \\
= &  \delta t  \left( \frac{\pi}{2} \coth \frac{\pi\omega_0\delta t}{2}\right)^{-1} \left(-\gamma_E - \Re  \psi \left(\frac{i}{2}\omega_0\delta t \right)+\frac{\tau}{\delta t} \right)^2 +  \delta t  \left( \frac{\pi}{2} \coth \frac{\pi\omega_0\delta t}{2}\right).\hfill \\
\end{split}
\label{a35}
\end{equation}
One special case of $\rho^2(\tau)$ is when $\omega_0 \delta t \ll 1$, in which $\frac{\pi}{2} \coth \frac{\pi\omega_0\delta t}{2} \sim \frac{1}{\omega_0 \delta t}$. Therefore,
\begin{equation}
\begin{split}
\rho^2(\tau) \sim &   \omega_0 \delta t^2 \left( -\frac{\zeta(3)}{4}\omega_0^2\delta t^2 +\frac{\tau}{\delta t} \right)^2 +    \frac{1}{\omega_0 } = \frac{1}{\omega_0} + \omega_0 \tau^2 + \mathcal{O}(\omega_0^3 \delta t^3).\hfill \\
\end{split}
\tag{\ref{a-8}}
\end{equation}
according to identity (\ref{a-7}). Another case is when $\omega_0 \delta t \gg 1$, in which $\coth \frac{\pi \omega_0 \delta t}{2} \to 1$. Thus by using the identity
\begin{equation}
\Re \psi(iy) \approx \log y + \sum_{n=1}^{\infty} \frac{(-1)^{n-1}B_{2n}}{2n y^{2n}} \sim \log y + \cO (y^{-2}),~ y \to \infty
\end{equation}
we can find 
\begin{equation}
\rho^2(\tau) \sim \delta t \left[ \frac{2}{\pi} \left(- \log \omega_0 \delta t + \log 2-\gamma_E +\frac{\tau}{\delta t} \right)^2  + \frac{\pi}{2}  \right]
\tag{\ref{a-9}}
\end{equation}

\section{Entanglement Entropy} \label{appB}

In this appendix we explicitly derive the approximated Entanglement Entropy (\ref{5-9}) and (\ref{5-11}). In appendix \ref{B-1} we figure out $\langle N_A \rangle$ and in appendix \ref{B-2}, we figure out $\int_{A_P \times A_P} {\text d} x {\text d} y |C(x,y) |^2$.

\subsection{$\langle N_A \rangle$}
\label{B-1}

First, we rewrite $\langle N_A \rangle$ into 
\begin{equation}
\langle N_A \rangle 
=   {\frac {1}{\Gamma(N)2^{N}\sqrt{\pi}}} \int_{A_{P} \times A_{P} } {\text d} \xi {\text d} \eta \delta(\xi-\eta)  e^{-\frac{\xi^2+\eta^2}{2}}{\frac {H_{N-1}(\eta)H_{N}(\xi)-H_{N-1}(\xi)H_{N}(\eta)}{\xi-\eta}} 
\label{na_ex}
\end{equation}
s.t. $\langle N_A \rangle$ has similar form to $\int_{A_P \times A_P} {\text d} x {\text d} y |C(x,y) |^2$ in (\ref{cxy_ex}). One can easily prove that (\ref{na_ex}) and (\ref{cxy_ex}) are identical.

In the large $N$ limit, the Hermite polynomial shows the following asymptotic behavior
\begin{equation}
e^{-{\frac {x^{2}}{2}}}\cdot H_{n}(x)\sim {\frac {2^{n}}{\sqrt {\pi }}}\Gamma \left({\frac {n+1}{2}}\right)\cos \left(x{\sqrt {2n}}-{\frac {n\pi }{2}}\right)
\label{b-1}
\end{equation}
We use this to simplify the integrand on the RHS of (\ref{cxy_ex}) or (\ref{na_ex}),
\begin{equation}
\begin{split}
 {\frac {1}{\Gamma(N)2^{N}\sqrt{\pi}}} e^{-\frac{\xi^2+\eta^2}{2}} & \left[ H_{N-1}(\eta)H_{N}(\xi)-H_{N-1}(\xi)H_{N}(\eta)\right] \hfill \\
= {\frac {1}{2\pi }} & \left[ -(-1)^N \sin \left(\eta{\sqrt {2N-2}}+ \xi{\sqrt {2N}}\right)- \sin \left(\eta{\sqrt {2N-2}}-\xi{\sqrt {2N}}\right) \right. \hfill \\
& \left. + (-1)^N \sin \left(\xi{\sqrt {2N-2}}+ \eta{\sqrt {2N}} \right)+ \sin \left(\xi{\sqrt {2N-2}}-\eta{\sqrt {2N}}\right) \right]. \hfill \\
\end{split}
\end{equation}

Now, change the variables of integration by defining
\begin{equation}
u \equiv \frac{\xi +\eta}{\sqrt{2}}, v \equiv \frac{\xi - \eta}{\sqrt{2}},
\end{equation}
and we obtain
\begin{equation}
\begin{split}
\langle N_A \rangle 
=  {\frac {1}{2\pi }} \int_{A_{P} \times A_{P} } {\text d} u {\text d} v   
{\frac {\delta(v)}{v}} &
 \left\{ -(-1)^N \cos \left[ (\sqrt{N-1}+\sqrt{N})u \right] \sin \left[  \frac{v}{\sqrt{N-1}+\sqrt{N}} \right]\right. \hfill \\
& \left. + \cos \left[ - \frac{u}{\sqrt{N-1}+\sqrt{N}}\right] \sin \left[ (\sqrt{N-1}+\sqrt{N})v \right] \right\}. \hfill \\
\end{split}
\end{equation}
Note that
\begin{equation}
\int_{A_P \times A_P} {\text d}u {\text d}v= \int_{0}^{\sqrt{2}\frac{a}{\rho}} {\text d} u \int_{|v|\le \sqrt{2}\frac{a}{\rho}-u} {\text d} v + \int_{-\sqrt{2}\frac{a}{\rho}}^{0} {\text d} u \int_{|v|\le \sqrt{2}\frac{a}{\rho}+u} {\text d} v 
= 2 \int_{0}^{\sqrt{2}\frac{a}{\rho}} {\text d} u \int_{|v|\le \sqrt{2}\frac{a}{\rho}-u} {\text d} v 
\end{equation}
since the integrand is even for both $u$ and $v$. Moreover, because of the Dirac delta function,
\begin{equation}
\int_{A_P \times A_P} {\text d}u {\text d}v \to 2 \int_{0}^{\sqrt{2}\frac{a}{\rho}} {\text d} u \int_{-\epsilon}^{\epsilon} {\text d} v, 
\end{equation}
and the integrand can be expanded around $v=0$:
\begin{equation}
\begin{split}
\langle N_A \rangle
=&  {\frac {1}{\pi }} \int_{0}^{\sqrt{2}\frac{a}{\rho}} {\text d} u \int_{-\varepsilon}^{\varepsilon} {\text d} v   {\delta(v)} 
 \left\{ -(-1)^N \cos \left[ (\sqrt{N-1}+\sqrt{N})u \right]   \frac{1}{\sqrt{N-1}+\sqrt{N}} \right. \hfill \\
& \left. + \cos \left[ - \frac{u}{\sqrt{N-1}+\sqrt{N}}\right]  (\sqrt{N-1}+\sqrt{N})  \right\} \hfill \\
=&  {\frac {1}{\pi }}  
 \left\{ (-1)^{N-1} \sin \left[ (\sqrt{N-1}+\sqrt{N})\sqrt{2}\frac{a}{\rho} \right]   \frac{1}{(\sqrt{N-1}+\sqrt{N})^2} + \sin \left[  \frac{\sqrt{2}\frac{a}{\rho}}{\sqrt{N-1}+\sqrt{N}}\right]  (\sqrt{N-1}+\sqrt{N})^2  \right\} \hfill \\
\to &  \frac{1}{\pi}(\sqrt{N-1}+\sqrt{N})\sqrt{2}\frac{a}{\rho} + \mathcal{O}\left(\frac{1}{N} \right)
\end{split}
\label{b-2}
\end{equation}
when 
\begin{equation}
\frac{\sqrt{2}\frac{a}{\rho}}{\sqrt{N-1}+\sqrt{N}} \ll 1.
\tag{\ref{5-7}}
\end{equation}

\subsection{$\int_{A_P \times A_P} {\text d} x {\text d} y |C(x,y) |^2$}
\label{B-2}

Similar to $\langle N_A \rangle$ (appendix \ref{B-1}),
\begin{equation}
\begin{split}
 \int_{A_P \times A_P} {\text d} x {\text d} y |C(x,y) |^2  \hfill~~~~~~~~ & \\
\approx  {\frac {2}{\pi^2 }}   \int_{0}^{\sqrt{2}\frac{a}{\rho}} {\text d} v \int_{0}^{\sqrt{2}\frac{a}{\rho}-v} {\text d} u  
& {\frac {1}{v^2}} 
 \left\{ -(-1)^N \cos \left[ (\sqrt{N-1}+\sqrt{N})u \right] \sin \left[  \frac{v}{\sqrt{N-1}+\sqrt{N}} \right]\right. \hfill \\
& \left. + \cos \left[  \frac{u}{\sqrt{N-1}+\sqrt{N}}\right] \sin \left[ (\sqrt{N-1}+\sqrt{N})v \right] \right\}^2. \hfill \\
\end{split}
\label{b-3}
\end{equation}

In the limit (\ref{5-7}), 
\begin{equation}
\sin \left[  \frac{v}{\sqrt{N-1}+\sqrt{N}} \right] \sim  \frac{v}{\sqrt{N-1}+\sqrt{N}} \ll 1 \sim \cos \left[  \frac{u}{\sqrt{N-1}+\sqrt{N}}\right].
\end{equation}
This implies that we can ignore the $1^{st}$ term in the integrand of (\ref{b-3}) and replace the cosine by 1. As a result,
\begin{equation}
\begin{split}
& \int_{A_P \times A_P} {\text d} x {\text d} y |C(x,y) |^2 
\to  {\frac {2}{\pi^2 }}   \int_{0}^{\sqrt{2}\frac{a}{\rho}} {\text d} v   
{\frac {1}{v^2}} 
 \sin^2 \left[ (\sqrt{N-1}+\sqrt{N})v \right] \left( \sqrt{2}\frac{a}{\rho}-v \right) \hfill \\
&\quad= -\frac{1}{\pi^2} \left\{ 1+\gamma_E - \cos \left[ (\sqrt{N-1}+\sqrt{N}) 2\sqrt{2}\frac{a}{\rho} \right] - \text{Ci}\left[ (\sqrt{N-1}+\sqrt{N}) 2\sqrt{2}\frac{a}{\rho} \right] \right. \hfill \\
 & \left.\qquad +\log \left[ (\sqrt{N-1}+\sqrt{N}) 2\sqrt{2}\frac{a}{\rho} \right] - \left[ (\sqrt{N-1}+\sqrt{N}) 2\sqrt{2}\frac{a}{\rho} \right] \text{Si}\left[ (\sqrt{N-1}+\sqrt{N}) 2\sqrt{2}\frac{a}{\rho} \right]\right\}
\end{split}
\label{b-4}
\end{equation}

The asymptotic behaviors of Trigonometric integrals are
\begin{equation}
\begin{split}
\operatorname{Si} (x)={\frac {\pi }{2}}-{\frac {\cos x}{x}}\left(1-{\frac {2!}{x^{2}}}+{\frac {4!}{x^{4}}}-{\frac {6!}{x^{6}}}\cdots \right)-{\frac {\sin x}{x}}\left({\frac {1}{x}}-{\frac {3!}{x^{3}}}+{\frac {5!}{x^{5}}}-{\frac {7!}{x^{7}}}\cdots \right) \hfill \\
\operatorname {Ci} (x)={\frac {\sin x}{x}}\left(1-{\frac {2!}{x^{2}}}+{\frac {4!}{x^{4}}}-{\frac {6!}{x^{6}}}\cdots \right)-{\frac {\cos x}{x}}\left({\frac {1}{x}}-{\frac {3!}{x^{3}}}+{\frac {5!}{x^{5}}}-{\frac {7!}{x^{7}}}\cdots \right)
\end{split}
\end{equation} 
when $x \to \infty$, and
\begin{equation}
\begin{split}
\operatorname {Si} (x)=\sum _{n=0}^{\infty }{\frac {(-1)^{n}x^{2n+1}}{(2n+1)(2n+1)!}}=x-{\frac {x^{3}}{3!\cdot 3}}+{\frac {x^{5}}{5!\cdot 5}}-{\frac {x^{7}}{7!\cdot 7}}\pm \cdots \\
\operatorname {Ci} (x)=\gamma_E +\ln x+\sum _{n=1}^{\infty }{\frac {(-1)^{n}x^{2n}}{2n(2n)!}}=\gamma_E +\ln x-{\frac {x^{2}}{2!\cdot 2}}+{\frac {x^{4}}{4!\cdot 4}}\mp \cdots 
\end{split}
\end{equation}
when $x \ll 1$. Thus (\ref{b-4}) can be further simplified into
\begin{equation}
\begin{split}
& \int_{A_P \times A_P} {\text d} x {\text d} y |C(x,y) |^2 
\to   -\frac{1}{\pi^2} \left\{ 1+\gamma_E  
 +\log \left[ (\sqrt{N-1}+\sqrt{N}) 2\sqrt{2}\frac{a}{\rho} \right] \right\}
 + \frac{1}{\pi} (\sqrt{N-1}+\sqrt{N}) \sqrt{2}\frac{a}{\rho},
\end{split}
\label{b-5}
\end{equation}
when 
\begin{equation}
\sqrt{N-1}+\sqrt{N} \gg \frac{a}{\rho} \gg \sqrt{N}-\sqrt{N-1};
\tag{\ref{5-8}}
\end{equation}
and
\begin{equation}
\begin{split}
& \int_{A_P \times A_P} {\text d} x {\text d} y |C(x,y) |^2
\to 0+ \mathcal{O}\left( \frac{Na^2}{\rho^2} \right) ,
\end{split}
\label{b-6}
\end{equation}
when 
\begin{equation}
 (\sqrt{N-1}+\sqrt{N})\frac{a}{\rho} \ll 1.
 \tag{\ref{5-10}}
\end{equation}
Inserting (\ref{b-5}) and (\ref{b-6}) back into (\ref{2-9}) together with (\ref{b-2}), one can get (\ref{5-9}) and (\ref{5-11}), respectively.

\end{document}